\documentclass{aa}
\usepackage{graphicx}
\usepackage{txfonts}

\begin{document} 

\def\beginthetable{ \begin{table*} }
\def\endthetable{ \end{table*} }

\def\absPath{"./"}
	
   \title{The problematically short superwind of OH/IR stars}
   \subtitle{Probing the outflow with the 69 $\mu$m spectral band of forsterite}

\author{B.L. de Vries \inst{1}
	\and J.A.D.L. Blommaert \inst{1,2}
	\and L.B.F.M. Waters \inst{3,4} 
	\and C. Waelkens \inst{1}
	\and M. Min \inst{4}
	\and R.~Lombaert \inst{1}
	\and H.~Van~Winckel \inst{1}
}

\authorrunning{B.L. de Vries et al.}

\institute{ Instituut voor Sterrenkunde, K.U. Leuven, Celestijnenlaan 200D, 3001 Leuven, Belgium
	\and Department of Physics and Astrophysics, Vrije Universiteit Brussel, Pleinlaan 2, 1050 Brussels, Belgium
	\and SRON Netherlands Institute for Space Research, P.O. Box 800, 9700 AV Groningen, The Netherlands
	\and Sterrenkundig Instituut Anton Pannekoek, University of Amsterdam, Science Park 904, 1098 XH, Amsterdam, The Netherlands
}

   \date{Received 27/08/2013; accepted 31/10/2013}
 
  \abstract
   {}
  {Spectra of OH/IR stars show prominent spectral bands of crystalline olivine ($\text{Mg}_{(2-2x)}\text{Fe}_{(2x)}\text{SiO}_{4}$). To learn more about the timescale of the outflows of OH/IR stars, we study the spectral band of crystalline olivine at 69~$\mu$m.}
   { The 69~$\mu$m band is of interest because its width and peak wavelength position are sensitive to the grain temperature and to the exact composition of the crystalline olivine. With Herschel/PACS, we observed the 69~$\mu$m band in the outflow of 14 OH/IR stars. By comparing the crystalline olivine features of our sample with those of model spectra, we determined the size of the outflow and its crystalline olivine abundance.}
   {The temperature indicated by the observed 69 $\mu$m bands can only be reproduced by models with a geometrically compact superwind ($R_{\rm{SW}}\lesssim$ 2500 AU = 1400 R$_{*}$). This means that the superwind started less than 1200 years ago (assuming an outflow velocity of 10 km/s). The small amount of mass lost in one superwind and the high progenitor mass of the OH/IR stars introduce a mass loss and thus evolutionary problem for these objects, which has not yet been understood.}
   {}
					

   \maketitle
%

\section{Introduction}
Although still far from a complete theory, the mass loss of asymptotic giant branch (AGB) stars is thought to be triggered by two processes: large-amplitude pulsations and radiation pressure on condensed dust grains \citep{wood79, castor81, holzer85, bowen88, hearn90}. Based on stellar evolution models and empirically determined mass loss rates, \cite{vassi93} show that typical AGB stars go through several superwind phases of increased mass loss in which they lose several tenths of a solar mass. 

Since the time when \cite{renzini81} introduced the idea of a superwind, many AGB stars were found to have mass loss rates consistent with the suggested superwind values \citep{knapp85, bedijn87, wood92}. The spatial extent of the superwind around oxygen-rich, high mass loss rate OH/IR stars is found to be extremely small. \cite{heske90} showed that the mass loss rates derived from CO are lower than those derived from infrared fluxes.  And detailed modeling of a sample of stars by \cite{just06} and \cite{just13} also showed that the mass loss rates derived from SED fitting and by solving the equation of motion for the dust drag wind, are higher than the mass loss rates derived from OH and CO observations. Since the OH and CO emission is from the outer parts of the wind and the dust emission is dominated by more recent mass loss, the authors conclude that a high density superwind  indeed exists that has not yet reached the distances where OH and CO lines are formed. Therefore the superwind must have started $<$2000 years ago. In a detailed study of the wind of WX~Psc, \cite{decin07} show that the current superwind phase of WX~Psc extends to $\sim$50 AU. Modeling of OH~26.5+0.6 indicated a superwind outflow extending up to $\sim$500 AU \citep{just96, ches05, groenewegen12}. 

OH/IR stars are known to contain moderate abundances of crystalline dust species like crystalline olivine ($\text{Mg}_{(2-2x)}\text{Fe}_{(2x)}\text{SiO}_{4}$ with $x$ between zero and one, \citealt{waters96, devries10, jones12}). Theoretical studies predict that the formation of crystalline olivine dust is dependent on the gas density and thus the mass loss rate of the star \citep{tielens98,gailsedl99,sogawa99}. High densities are more favorable to the formation of crystals than low densities. Even though a temperature contrast makes it difficult to observe crystalline olivine in low mass loss rate AGB stars \citep{kemper01}, it is shown that crystalline olivine production increases with the mass loss rate of the AGB star \citep{jones12, speck08}. Since the superwind phases of OH/IR stars create a much denser environment, the formation of crystals is expected to be linked with characteristics of the superwind, making the study of crystalline olivine features in the spectra of OH/IR stars interesting.

	\begin{figure}
	\begin{center}
	\resizebox{\hsize}{!}{\includegraphics{\absPath 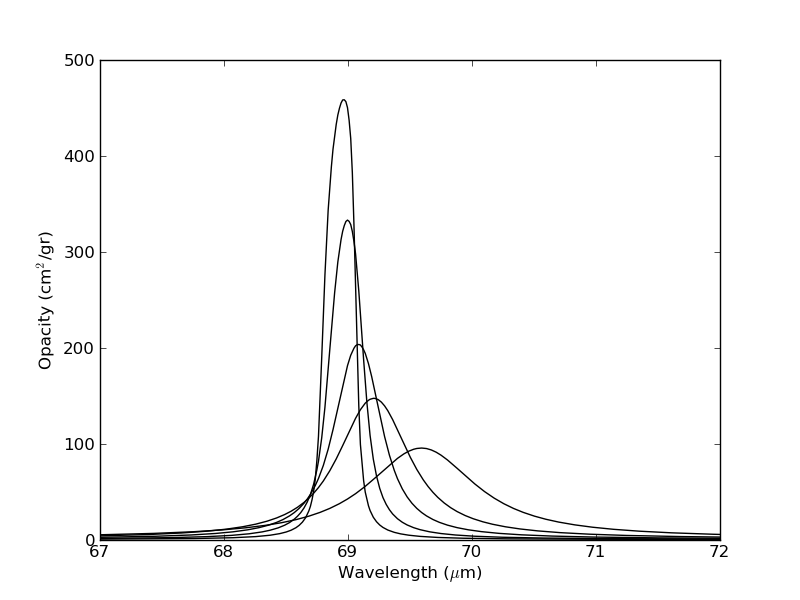}}
	\caption{Opacities of crystalline olivine in the region of the 69~$\mu$m band for five different temperatures (pure forsterite, \citealt{suto06}). From the strongest, bluest and most narrow band to the weakest, reddest and broadest band, the opacities are for crystalline olivine with a grain temperature of 50, 100, 150, 200 and 295~K. }
	\label{fig: opacs_T}
	\end{center}
	\end{figure}

In this work we introduce a new diagnostic to probe the superwind of OH/IR stars based on the 69~$\mu$m spectral band of crystalline olivine. With Herschel/PACS we observed the 69~$\mu$m band of crystalline olivine with great sensitivity. This band is of great interest because its width and peak wavelength position are sensitive to the grain temperature and the exact composition of the crystalline olivine \cite{koike03, suto06}, see fig. \ref{fig: opacs_T}). Because of this temperature dependence the 69~$\mu$m band can be used to probe the spatial distribution of the crystalline olivine and thus of the outflow.

In this work we present 14 OH/IR stars which were observed with Herschel/PACS in the wavelength range of the 69~$\mu$m band of crystalline olivine (Blommaert et al in prep). We first introduce the sample of stars in sect. \ref{sec: observations}, describe the spectral features of crystalline olivine in sect. \ref{sec: feats} and list the methods used in sect. \ref{sec: method}. The modeling is presented in sect. \ref{sec: models}, followed by the results in sect. \ref{sec: generalresults}. We finish with a discussion and several conclusions in sect. \ref{sec: discussion}.

\beginthetable
\centering
\begin{tabular}{ l l l l l l l l l }

  Source name	&	IRAS			&	\# 	&	\.{M} ($10^{-5}\,\text{M}_{\odot}/\text{yr}$)	&11.3~$\mu$m&33.6 $\mu$m&	69.0 $\mu$m	&	P(days)	&	L(L$_{\odot}$)	\\
  \hline \hline
 OH~32.8-0.3  	&	18498$-$0017	&	1	&	16 (1)					&+			&+						&	+		&	1,539		&	23,100\\
 OH~21.5+0.5 	&	18257$-$1000	&	2	&	26 (2)					&?			&+						&	+		&	1,785		&	72,100-96,900\\
 OH~30.1-0.7  	&	18460$-$0254	&	3	&	18 (3)					&+			&+						&	+		&	2,013		&	-	\\
 OH~26.5+0.6  	&	18348$-$0526	&	4	&	3.2 (4)					&+			&+						&	+		&	1,570		&	39,362\\
 OH~127.8+0.0	&	01304+6211	&	5	&	9.2 (3)					&-			&+						&	+		&	1,525		&	38,012\\
 AFGL~2403  	&	19283+1944	&	6	&	1.7 (5)					&+			&+						&	+		&	-			&	1,574\\
 IRAS~21554  	&	21554+6204	&	7	&	1.5 (5)					&?			&+						&	+		&	-			&	4,984\\
 AFGL~5379  	&	17411$-$3154	&	8	&	9.8 (3)					&-			&+						&	+		&	1,440		&	35,484\\
 OH~104.91+2.4&	22177+5936	&	9	&	5.6 (6)					&-			&+						&	-		&	1,620		&	40,871\\
 AFGL~4259  	&	20043+2653	&	10	&	-						&-			&+						&	-		&	-			&	-\\
 IRAS~17010  	&	17010$-$3840	&	11	&	-						&+			&+						&	-		&	-			&	-\\
 RAFGL~2374 	&	19192+0922	&	12	&	0.46 (5)					&?			&+						&	-		&	534			&	4,300\\
 WX~Psc  		&	01037+1219	&	13	&	1.7 (3)					&?			&+						&	-		&	660			&	13,914\\
 IRC+50137 	&	05073+52  48	&	14	&	0.6 (7)					&?			&+						&	-		&	635			&	13,284\\
\hline	
\end{tabular}
 \caption{ List of the observed OH/IR stars and possible detections of crystalline olivine spectral bands. References for the mass loss rates: (1) \cite{groene94}, (2) \cite{just06}, (3) \cite{just13}, (4) \cite{just96}, (5) \cite{beck10}, (6) \cite{riechers05}, (7) \cite{just92}. References for the periods and luminosities are \cite{vanlangevelde90} and \cite{beck10} and references therein. 
 }
\label{tab: tableSources}
\endthetable

	\begin{figure}
	\begin{center}
	\resizebox{\hsize}{!}{\includegraphics{\absPath 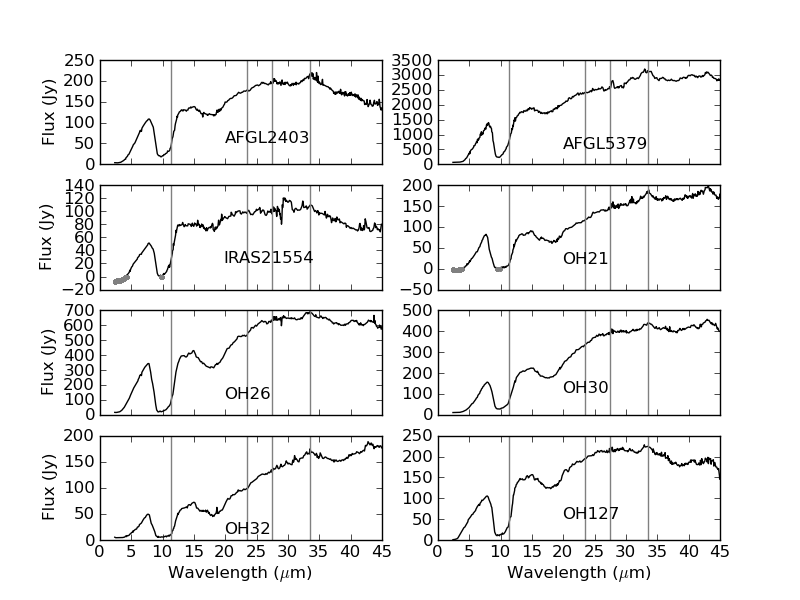}}
	\caption{In black ISO-SWS spectra for the OH/IR stars. The data are taken from \cite{sloan03}. For the sources OH~21.5+0.5 and IRAS~21554 the flux levels go below zero in regions below 5 $\mu$m and in the 9.7 $\mu$m band. Those negative flux values are indicated with gray dots. Peaks in the opacity curve of crystalline olivine at 11.3, 23.5, 27.5 and 33.5 $\mu$m are indicated with vertical g lines.}
	\label{fig: ISO_spec_det}
	\end{center}
	\end{figure}

	\begin{figure}
	\begin{center}
	\resizebox{\hsize}{!}{\includegraphics{\absPath 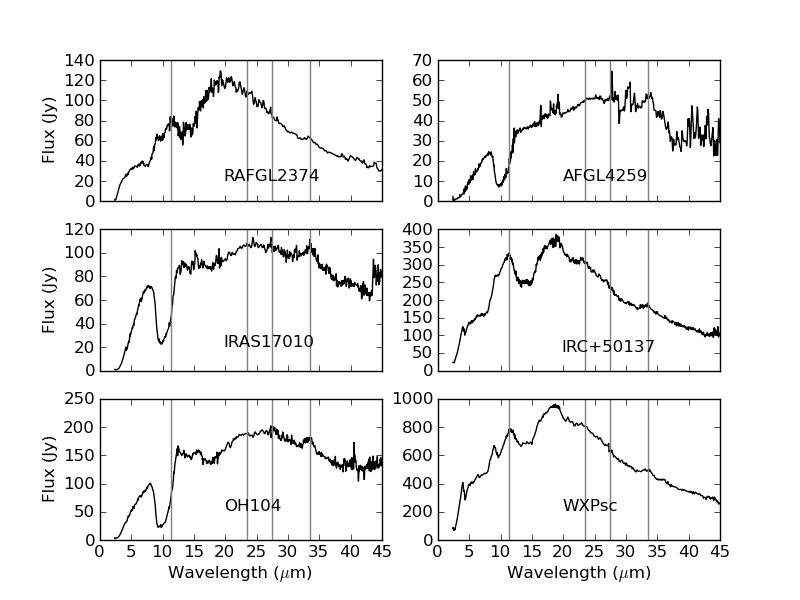}}
	\caption{Continuation of fig. \ref{fig: ISO_spec_det} }
	\label{fig: ISO_spec_ndet}
	\end{center}
	\end{figure}

\section{Sample selection and data reduction}
\label{sec: observations}
We selected 14 OH/IR stars based on the presence of crystalline olivine bands in their ISO-SWS spectrum (see fig. \ref{fig: ISO_spec_det} and \ref{fig: ISO_spec_ndet} and table \ref{tab: tableSources}). Crystalline olivine can be best recognized in the ISO-SWS spectra by its emission band at 33.6 $\mu$m or by the absorption feature at 11.3~$\mu$m on the red wing of the 9.7 $\mu$m band of amorphous silicate. See fig. \ref{fig: opacs} for the opacities of crystalline olivine in the ISO-SWS wavelength range. The central star of the objects in our sample are Galactic disk sources with luminosities going up to several $10,000\,\text{L}_{\odot}$ (see table \ref{tab: tableSources}, \citealt{vanlangevelde90}). The central star of the objects is completely obscured by their circumstellar dust shell and their infrared spectrum shows an 9.7 $\mu$m absorption band of amorphous silicate. Only WX~Psc, RAFGL~2374 and IRC~+50137 have their 9.7 $\mu$m bands in-between an emission and absorption feature. The infrared spectra of these sources range from relatively blue spectra combined with weak 9.7 $\mu$m absorption or emission bands to relatively red spectra with very deep  9.7 $\mu$m bands. The blue and red spectra correspond to relatively low ($10^{-6}\,$M$_{\odot}/$yr) and high ($10^{-4}\,$M$_{\odot}/$yr) mass loss rates respectively. In table \ref{tab: tableSources} the sources are ordered from reddest to bluest from top to bottom. 

The observations were taken with the spectrometer Herschel/PACS \citep{pilbratt10, poglitsch10}. The objects have been previously published by Blommaert et al. (in prep.) as part of a large sample of observations of the 69~$\mu$m band in the spectra of all kinds of evolved stars. For the sources  OH~32.8-0.3, OH~30.1-0.7, OH~104.91+2.4 and IRAS~17010 we have spectra in the wavelength range 67-72 $\mu$m, which are from the Guarenteed time program "Forsterite dust in the circumstellar environment of evolved stars" (GT1\_jblommae\_1, Blommaert et al in prep). For the sources OH~21.5+0.5, OH~26.5+0.6, AFGL~2403, IRAS~21554, AFGL~4259, RAFGL~2374, IRC~+50137 and AFGL2019 we also have spectra in the wavelength range of 67-72 $\mu$m, which are from the open time program "Study of the cool forsterite dust around evolved stars" (OT2\_jblommae\_2, Blommaert et al in prep). For the sources OH~127.8+0.0, AFGL~5379 and WX~Psc we have full spectral energy distribution Herschel/PACS spectra (50-200~$\mu$m), where OH~127.8+0.0 is a PACS calibration measurement and the other two are from the key program MESS \citep{groenewegen11_mess}. 

The data reduction of the Herschel/PACS spectra is fully described in Blommaert et al. (in prep.), but is briefly repeated here. The spectra were reduced in the Herschel Interactive Processing Environment (HIPE, \citealt{ott10}) package version 9, using the Herschel/PACS pipe-line script (ChopNodRangeScan.py). The absolute flux calibration used the PACS internal calibration block measurements and for the spectral shape the PACS Relative Spectral Response Function was applied. The PACS integral-field-spectrometer contains 5$\times$5 spatial pixels, the so-called spaxels. The point spread function (PSF) of Herschel is larger than the central spaxel ($9.4^{\prime\prime} \times 9.4^{\prime\prime}$) and a correction needs to be made for the missing part of the PSF, even when combining several spaxels, to obtain the absolute source flux. Because of the pointing accuracy and jitter, it is best to combine as many spaxels under the PSF as possible. However, in order to keep the noise as low as possible, we combined all spaxels with a signal to noise (S/N) larger than 10 rather than blindly adding the central 3$\times$3 spaxels. The S/N of every spaxel is obtained by calculating the standard deviation from a linear fit to the wavelength region of 67.5 to 68.0 $\mu$m and dividing this standard deviation by the signal. 

The sources AFGL~5379 and OH~21.5+0.5 were pointed off-target. For OH~21.5+0.5 all of the flux was still contained within the 5$\times$5 spaxels and a spectrum could be extracted. The maximum flux within the spaxels for AFGL~5379 was found on one of the border spaxels, and thus not all flux of this object was detected and therefor the flux levels should be taken as a lower limits. We flux calibrated the spectra by calculating the total continuum flux (in the range of 67.5 to 68.0 $\mu$m) in the spectrum of all 3$\times$3 spaxels combined and corrected this for the PSF (the flux levels were multiplied with a correction factor of 1.09, see the PACS data reduction manual). This flux value was used to scale the combined spaxels with S/N$>$10 to the proper continuum flux level.  

The ISO-SWS spectra were taken from the Sloan database \citep{sloan03}. The ISO-SWS spectra of the sources IRAS~21554 and OH~21.5+0.5 go below zero at wavelengths smaller than 5 $\mu$m and in the 9.7 $\mu$m band of amorphous silicate. This is probably due to uncertainties in the dark current of the detector (see fig. \ref{fig: ISO_spec_det}).

\section{Crystalline olivine spectral features}
\label{sec: feats}
\begin{figure}
	\begin{center}
	\resizebox{\hsize}{!}{\includegraphics{\absPath 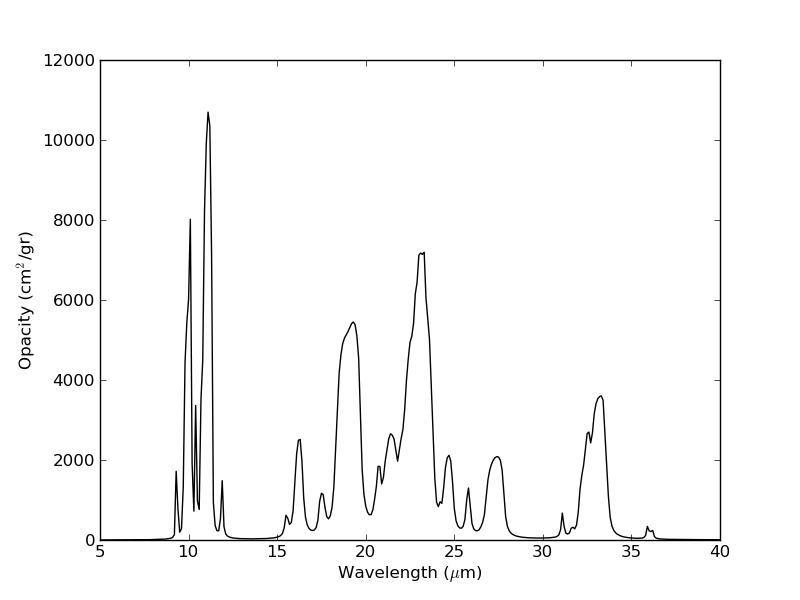}}
	\caption{Opacities of crystalline olivine in the 5-40 $\mu$m wavelength range. The opacities are for crystalline olivine with a grain temperature of 150~K and calculated with CDE particle shapes (see sect. \ref{sec: radtransport}). The optical constants are from \cite{suto06}.}
	\label{fig: opacs}
	\end{center}
	\end{figure}

In this section we describe three key spectral features of crystalline olivine: the 11.3, the 33.6 and the 69.0 $\mu$m bands. Figure~\ref{fig: opacs} shows the opacities of crystalline olivine in the 5-40 $\mu$m range. For a typical OH/IR star the 11.3~$\mu$m band is in absorption on top of the red wing of the 9.7 $\mu$m band of amorphous silicate and the 33.6 $\mu$m and 69~$\mu$m bands are in emission. Between the 11.3 and 33.6~$\mu$m bands crystalline olivine also has strong resonances at 23.5 and 27.5~$\mu$m, but in the spectra of OH/IR stars these bands are in self-reversal and difficult to discern in the spectra. (see fig. \ref{fig: ISO_spec_det} and \ref{fig: ISO_spec_ndet}).

\subsection*{\small{The 11.3~$\mu$m band}}
\label{sec: the113band}
The 11.3~$\mu$m band is discussed in \cite{devries10} and they show that when the optical depth of the dust shell is high at these wavelengths, the strength of this band is only dependent on the crystalline olivine abundance and not on other properties of the dust shell or central star. The 11.3~$\mu$m band probes the material close to the central star and in the line of sight to the central star. The presence of a 11.3~$\mu$m band can be best seen in plots of the optical depth shown in figures \ref{fig: ISO_113_det} and \ref{fig: ISO_113_ndet}. For the three stars with 9.7~$\mu$m bands that are in-between an emission and absorption band (RAFGL~2374, WX~Psc, IRC~+50137) we could not measure the 11.3~$\mu$m band, indicated with a "?" in table \ref{tab: tableSources}. The sources OH~21.5+0.5 and IRAS~21554 have ISO-SWS spectra that go below zero in the 9.7 $\mu$m band of amorphous olivine and because of this we do not analyze the 11.3~$\mu$m bands of these sources (also indicated with a "?" in table \ref{tab: tableSources}). Among the sources that have a detected 69~$\mu$m band we find that  OH~32.8-0.3, OH~30.1-0.7, OH~26.5+0.6 and AFGL~2403 also show a 11.3~$\mu$m band. For the objects OH~127.8+0.0 and AFGL~5379 we do not detect an 11.3~$\mu$m band while they do show a 69~$\mu$m band. Among the sources without detected 69~$\mu$m bands we only find an 11.3~$\mu$m band for IRAS~17010. 

	\begin{figure}
	\begin{center}
	\resizebox{\hsize}{!}{\includegraphics{\absPath 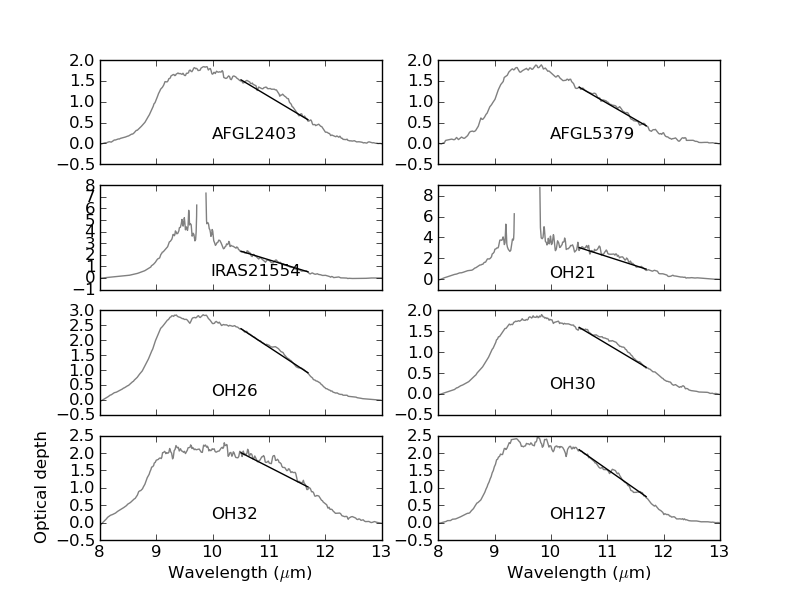}}
	\caption{ In gray the optical depth in the region of the 9.7 $\mu$m band of amorphous silicate. The calculation of the optical depth is described in sect. \ref{sec: measuring} and \cite{devries10}. In black the local continuum used under the 11.3~$\mu$m band. The flux levels of IRAS~21554 and OH~21.5+0.5 go below zero in the region of 9.7 $\mu$m. Those points are not used to calculate the optical depth and are excluded from these plots.}
	\label{fig: ISO_113_det}
	\end{center}
	\end{figure} 
	
	\begin{figure}
	\begin{center}
	\resizebox{\hsize}{!}{\includegraphics{\absPath 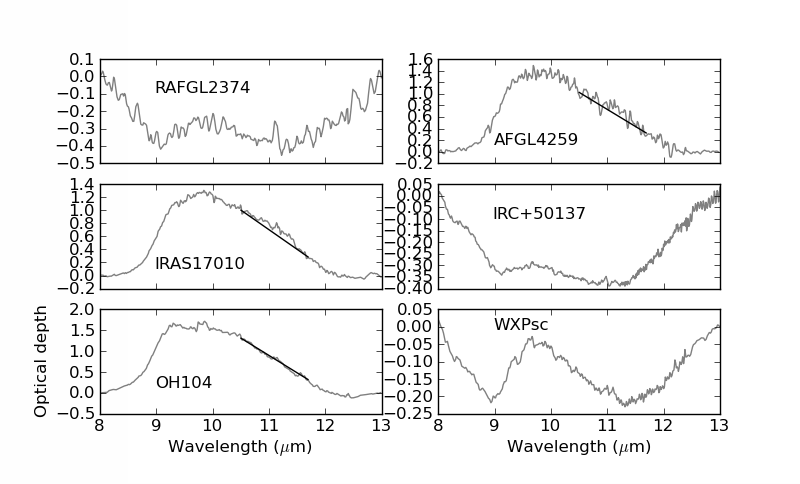}}
	\caption{Continuation of fig. \ref{fig: ISO_113_det}}
	\label{fig: ISO_113_ndet}
	\end{center}
	\end{figure}

\subsection*{\small{The 33.6 $\mu$m band}}
Figure \ref{fig: ISO_336_DET} and \ref{fig: ISO_336_nDET} show the continuum divided 33.6 $\mu$m bands of the 14 OH/IR stars. Because the band is easily recognized in the spectra and was used to select the sample, we considered using it for our quantitative analysis. However, we found this is not trivial. First because many of these bands show a double peaked profile for the 33.6 $\mu$m band, indicating it is probably a blend of more than one spectral band. Which other dust species is present in this spectral feature is not clear. And secondly, \cite{devries10} showed that the observed strength of the 33.6 $\mu$m band in OH/IR type stars is dependent on many of the properties of the dust shell, since the optical depth at this wavelength is already larger than one. In this study we use the 33.6 $\mu$m band to determine the presence of crystalline olivine, but because the band has a complicated structure and is strongly dependent on the outflow parameters, we do not use this band quantitatively in any further analysis.

	\begin{figure}
	\begin{center}
	\resizebox{\hsize}{!}{\includegraphics{\absPath 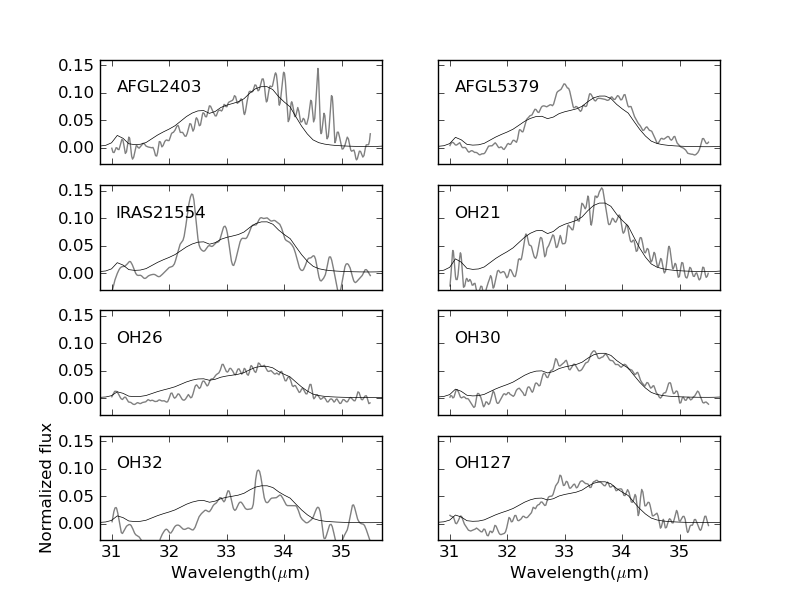}}
	\caption{Normalized 33.6 $\mu$m bands derived from the spectra in the Sloan database \citep{sloan03}. The normalization is done by calculating $(F/F_{\text{continuum}})-1$. The scale of the y-axis is the same for all plots. In black are plotted the opacities of crystalline olivine (at 150 K, using CDE, see sect.~\ref{sec: radtransport}). The opacities are normalized to the level of the observed band in the range of 33.4-33.6 $\mu$m.}
	\label{fig: ISO_336_DET}
	\end{center}
	\end{figure} 

	\begin{figure}
	\begin{center}
	\resizebox{\hsize}{!}{\includegraphics{\absPath 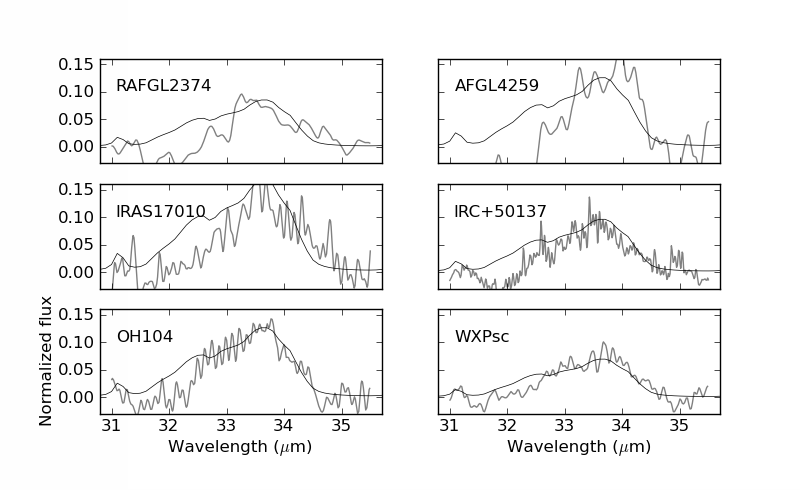}}
	\caption{Continuation of fig. \ref{fig: ISO_336_DET}.}
	\label{fig: ISO_336_nDET}
	\end{center}
	\end{figure} 
	
\subsection*{\small{The 69~$\mu$m band}}
For the reddest and most obscured sources in our sample, we detected the 69~$\mu$m band, but for the six bluest and lowest mass loss rate sources in our sample we did not (see fig. \ref{fig: PACS_spec_det} and \ref{fig: PACS_spec_ndet}). Already by eye it can be seen that the 69~$\mu$m band of IRAS~21554 is the narrowest and the one of AFGL~2403 the broadest.

The weakest, but still detected 69~$\mu$m bands, are seen in the spectrum of OH~26.5+0.6 and AFGL~5379, which also have weak and almost undetectable 11.3~$\mu$m bands. The sources  OH~32.8-0.3, AFGL~2403, OH~21.5+0.5 and IRAS~21554 show very strong 69~$\mu$m bands. For  OH~32.8-0.3 and AFGL~2403 we indeed also detect strong 11.3~$\mu$m bands. The 11.3~$\mu$m bands of OH~21.5+0.5 and IRAS~21554 are difficult to analyze because of their uncertain ISO-SWS flux levels. For the six sources that do not have detected 69~$\mu$m bands we also do not find 11.3~$\mu$m bands, or we can not see them because of the self reversal of the 9.7 $\mu$m band. Only IRAS~17010 has a convincing detection at 11.3~$\mu$m while having none at 69~$\mu$m. 

Most sources show gas lines in the Herschel/PACS range, which can be attributed to species like ortho- and para-H$_{2}$O, $^{12}$CO, $^{28}$SiO, and SO$_{2}$ (\citealt{decin12}, Blommaert et al in prep.). A blent of gas lines, although weak, is situated on top of the 69~$\mu$m band. There are only three sources in our sample for which we do not detect these gas lines: IRAS~21554, AFGL~4259 and IRAS~17010.

	\begin{figure}
	\begin{center}
	\resizebox{\hsize}{!}{\includegraphics{\absPath 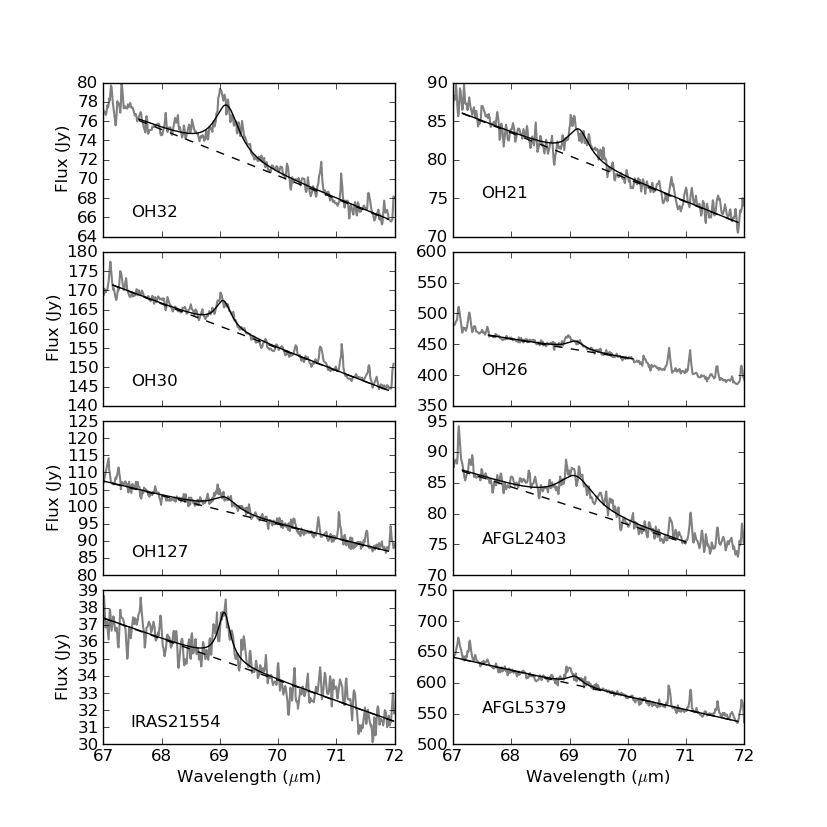}}
	\caption{In gray Herschel/PACS spectra for the OH/IR stars with 69~$\mu$m crystalline olivine band detections. In dashed and solid black the fitted continuum and fitted Lorentzian profile.}
	\label{fig: PACS_spec_det}
	\end{center}
	\end{figure} 
	
	\begin{figure}
	\begin{center}
	\resizebox{\hsize}{!}{\includegraphics{\absPath 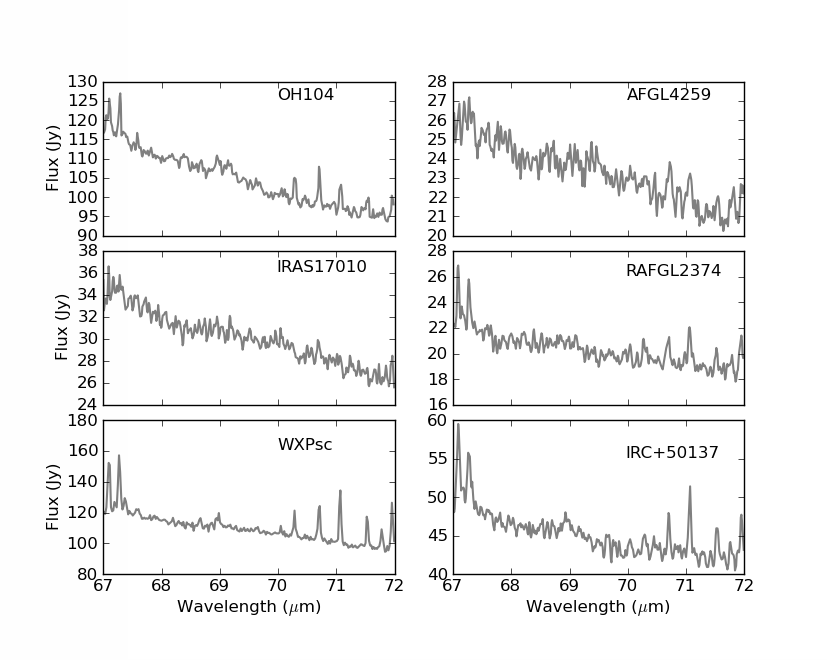}}
	\caption{Continuation of fig. \ref{fig: PACS_spec_det}}
	\label{fig: PACS_spec_ndet}
	\end{center}
	\end{figure} 
	
\section{Method and observational diagrams}
\label{sec: method}
In this study we want to answer specific questions regarding the distribution of crystalline olivine in the circumstellar environment of OH/IR stars. We can quantitatively determine how much and where the crystalline olivine is in the circumstellar environment by studying the two spectral features at 11.3 and 69~$\mu$m. In this section we discuss which properties of these bands are used, how we measure them and see how they relate to each other. In sect. \ref{sec: generalresults} we compare the properties of the observations with those of models. 

\subsection{Selection of spectral properties and motivation}

As a probe of the spatial distribution of the crystalline olivine material we take the width of the 69~$\mu$m band. The width of this band is sensitive to the grain temperature of the crystalline olivine and thus to the temperature gradient of crystalline olivine in the circumstellar environment. Using the models in sect. \ref{sec: models} we indeed show that the width of the 69~$\mu$m band can be used to determine up to what radius crystalline olivine is present in the circumstellar environment. The sources in our sample are all outflow sources and if the distribution of crystalline olivine is the same as the distribution of the other dust species, the width of the 69~$\mu$m band is a probe of the size of the outflow and also of the timescale of the outflow. The width of the 69~$\mu$m band could also depend on other parameters like the mass loss rate and we investigate these possible dependencies in sect. \ref{sec: models}. 

We also measure the peak-over-continuum strength of the 69~$\mu$m band. The strength of this band is, among others, dependent on the abundance of the crystalline olivine. Because the wavelength position of the band is at such long wavelengths, all the line of sights probe deep into the outflow and the abundance derived from the strength of the 69~$\mu$m band is representative of the average abundance in the whole outflow. This is in contrast to what \cite{devries10} have shown for the strength of the 11.3~$\mu$m band. \cite{devries10} define the strength of the 11.3~$\mu$m band as:
\begin{equation}
\label{eq: S113}
\text{S}_{11.3} = \int    \left( \frac{ \tau_{\text{cr}} }{ \tau_{\text{all}} } -1 \right)    \,\mathrm{d}\lambda
\end{equation}
Where the integral d$\lambda$ sums over the spectral feature and $\tau_{\text{cr}}$ and $\tau_{\text{all}}$ are the optical depth of crystalline olivine only and the optical depth of all dust species together, respectively. In the optical thick limit, equation \ref{eq: S113} can be approximated by:
\begin{equation}
\text{S}_{11.3}  = \int \int \left( \frac{ \rho(R) \times A_{\text{cr}} \times \kappa_{\text{fo}} }{ \rho(R) \times \kappa_{\text{all}} } - 1 \right)    \,\mathrm{d}V\,\mathrm{d}\lambda  \propto A_{\text{cr}}
\end{equation}
Where $\rho(R)$ is the density in the outflow, $A_{\text{cr}}$ the abundance of crystalline olivine and $\kappa_{\text{cr}}$ and $\kappa_{\text{all}}$ the opacity of only crystalline olivine and that of all dust species together, respectively. The integral d$\lambda$ sums over the spectral feature, while d$V$ sums over the volume of the outflow. In order to make the last step we assume that the crystalline olivine abundance is constant throughout the outflow. \cite{devries10} have indeed shown that if the optically thick limit holds, the strength of the 11.3~$\mu$m band is only dependent on the abundance of crystalline olivine in the line of sight. They show that the optically thick regime is valid when the mass loss rate is higher than $\sim5\cdot10^{-5}\,\text{M}_{\odot}/\text{yr}$ (with a gas to dust ratio of 100). This shows that in the optically thick regime, the strength of the 11.3~$\mu$m band is a good probe of the abundance of crystalline olivine in the line of sight and close to the central star. The combination of the strength of the 11.3~$\mu$m band and the strength of the 69~$\mu$m band are therefore a good probe of a spatial abundance gradient in the circumstellar material.

In this study we determine the distribution of crystalline olivine from its spectral features and do not aim to determine the properties of the amorphous dust or mass loss rates from the SED. But we still want to have a quantitative measure of the color of the SED in order to compare it with the properties of the 11.3 and 69~$\mu$m bands. In order to characterize the color of the SED we measure the flux ratio $F(30\,\mu \text{m})/F(18\,\mu \text{m})$. Several of our sources with the bluest SEDs of our sample have flux values with $F(30\,\mu \text{m})<F(18\,\mu \text{m})$ (for example WX~Psc or RAFGL~2374, see fig. \ref{fig: ISO_spec_det} and \ref{fig: ISO_spec_ndet}). And other stars have very red SEDs with $F(30\,\mu \text{m})>F(18\,\mu \text{m})$, like for example OH~26.5+0.6 or OH~21.5+0.5.

\subsection{Measuring the spectral properties}
\label{sec: measuring}

\subsubsection*{\small{Measuring the 11.3~$\mu$m band}}
We briefly describe how we measured the strength of the 11.3~$\mu$m band and the reader is referred to \cite{devries10} for the details. As shown in equation \ref{eq: S113}, the strength of the 11.3~$\mu$m band is calculated as the ratio of the optical depth due to only crystalline olivine ($\tau_{\text{cr}}$) and the optical depth of all dust species ($\tau_{\text{all}}$). We first construct a continuum over the 9.7~$\mu$m band of amorphous olivine. The optical depth in the 9.7~$\mu$m band is calculated using $\tau = -\text{ln}(F/F_{\text{cont}})$, where $F_{\text{cont}}$ is the flux level of the continuum constructed over the 9.7~$\mu$m band of amorphous silicate. Figure \ref{fig: ISO_113_det} and \ref{fig: ISO_113_ndet} show the optical depth in the 9.7~$\mu$m band for the sources in this study. The optical depth in the 9.7~$\mu$m band due to crystalline olivine ($\tau_{\text{cr}}$) and due to the other dust species ($\tau_{\text{all}}$) is now determined by constructing a continuum in $\tau$-space, under the 11.3~$\mu$m band (for examples see fig. \ref{fig: ISO_113_det} and \ref{fig: ISO_113_ndet}). The strength of the 11.3~$\mu$m band is now calculated by integrating over the 11.3~$\mu$m band as shown in equation \ref{eq: S113}.

The two sources OH~127.8+0.0 and AFGL~5379 have no 11.3~$\mu$m bands that we could identify by eye (see table \ref{tab: tableSources} and fig. \ref{fig: ISO_113_det}). But even though we do not recognize the feature, we applied the calculation of the 11.3~$\mu$m band to these sources and treat the measured values as upper limits for the strength of the 11.3~$\mu$m band, see sect. \ref{sec: obsDiag}.

\subsubsection*{\small{Measuring the 69~$\mu$m band}}
The observations are fitted over the range of 67-72 $\mu$m with a Lorentzian together with a slightly curved continuum (second order polynomial, see figures \ref{fig: PACS_spec_det} and \ref{fig: PACS_spec_ndet}). Most of the OH/IR stars show gas lines in their spectra and we excluded the following wavelength regions in our fits (all in $\mu$m): 66.9-67.15; 67.22-67.41; 67.47-67.6; 68.91-69.02; 70.6-70.8; 71.0-71.15; 71.5-71.6 and 71.9-72.2. 

The peak-over-continuum strength is obtained by dividing the amplitude of the Lorentzian fit by the continuum flux at the peak wavelength position of the feature.  For stars without a detection at 69~$\mu$m we determined an upper limit for the peak-over-continuum strength of the band. To do this we made a linear fit to the wavelength range 67.5-68.2 $\mu$m in the Herschel/PACS spectrum and took the standard deviation of the flux relative to this continuum as an upper limit of the peak-over-continuum strength of the 69~$\mu$m band. 

\subsubsection*{\small{Quantitative measure of the color of the spectrum}}
The fluxes at 18 and 30 $\mu$m for the flux ratio $F(30\,\mu \text{m})/F(18\,\mu \text{m})$ are calculated from the ISO-SWS spectra by taking the average in the regions 17.5-18.5 and 29.5-30.5~$\mu$m, respectively. 

\subsubsection*{\small{Error calculation}}
The errors on the width, position and strength of the 69~$\mu$m band and the strength of the 11.3~$\mu$m band are obtained by a Monte Carlo method. The error on the properties of the bands are calculated by repeating the analysis of the bands multiple times, while randomly varying (by $\pm$0.02 $\mu$m from the standard value) the wavelength location of the fitting boarders, the boarders of the continuum and the wavelength location of the gas lines that are excluded. The standard deviation of the sample of fitting results is than taken as the error. 

	\begin{figure}
	\begin{center}
	\resizebox{\hsize}{!}{\includegraphics{\absPath 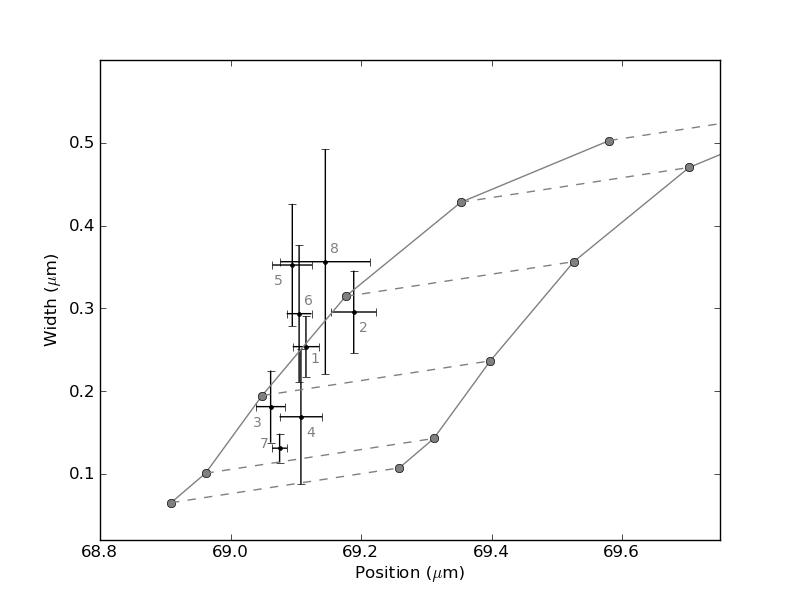}}
	\caption{The width and position of the observed 69~$\mu$m bands are plotted against each other in black dots. Gray solid and dashed lines are the width and position of 69~$\mu$m bands from laboratory measurements and interpolations between them at different temperatures and compositions. The left solid gray curve is for crystalline olivine with 0\% iron and with temperatures going from 50, 100, 150, 200, 295~K for the narrowest to broadest features.  The right solid gray curve is for the same temperatures but for crystalline olivine with 1\% iron (based on the interpolations of \citealt{devries12}). Dashed gray lines connect points with the same temperature.}
	\label{fig: PW_sourcesonly}
	\end{center}
	\end{figure}

	\begin{figure}
	\begin{center}
	\resizebox{\hsize}{!}{\includegraphics{\absPath 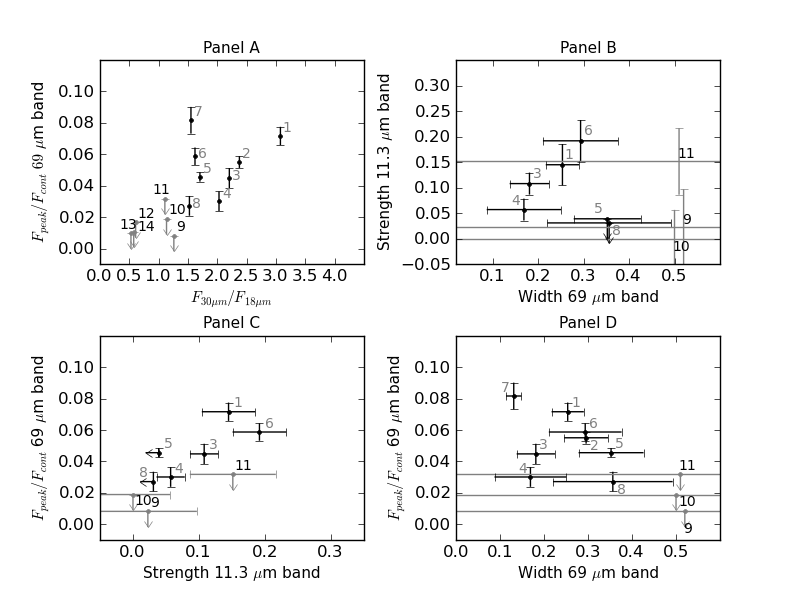}}
	\caption{The panels show different properties of the observed spectra. The observations are plotted in black and gray dots for the observations with and without detected 69~$\mu$m bands respectively. In panel A is plotted the peak-over-continuum strength of the 69~$\mu$m band versus the $F_{30\mu \text{m}}/F_{18\mu \text{m}}$ color of the spectrum. In panel B the strength of the 11.3~$\mu$m band is plotted against the width of the 69~$\mu$m band. In panel C the peak-over-continuum strength of the 69~$\mu$m band and the strength of the 11.3~$\mu$m band are shown. The peak-over-continuum strength of the 69 is plotted against the width of the 69~$\mu$m band in panel D.}
	\label{fig: diag_sourcesonly}
	\end{center}
	\end{figure} 

\subsection{Observational diagrams}
\label{sec: obsDiag}
We summarize the properties of the 11.3~$\mu$m and 69~$\mu$m bands and the color of the spectrum with five diagnostic diagrams. The first diagram shows the width plotted against the position of the 69~$\mu$m bands (see fig. \ref{fig: PW_sourcesonly}). The width and position of the laboratory measurements of the 69~$\mu$m bands shown in gray in fig.~\ref{fig: PW_sourcesonly} are based on the data sets of \cite{suto06} and \cite{koike03}. All details about how these curves are generated from the laboratory measurements are explained in \cite{devries12}. From fig. \ref{fig: PW_sourcesonly} it can be seen that the observations have positions and widths close to the curve of crystalline olivine with 0\% iron and with grain temperatures ranging from 100-200 K.

Figure \ref{fig: diag_sourcesonly}A shows the peak-over-continuum strength of the 69~$\mu$m band as a function of the color of the spectrum. Besides the source IRAS~21554 (\#7), all sources follow a diagonal trend where the 69~$\mu$m band becomes stronger as the spectrum becomes redder. This is an important diagram because it shows that the emission from the crystalline olivine is related to the overall emission of the dusty outflow.  

Panel B from fig. \ref{fig: diag_sourcesonly} shows the strength of the 11.3~$\mu$m band plotted against the width of the 69~$\mu$m band. Even though the sources seem to scatter over the diagram, this diagram is very useful in reading of the crystalline olivine grain temperature from the width of the 69~$\mu$m band and the crystalline olivine abundance in the line of sight from the 11.3~$\mu$m band. The diagram shows that the abundance of crystalline olivine in the line of sight does not correlate with the crystalline olivine grain temperature (which is a probe of the extent of the outflow, see sect. \ref{sec: models}).

In fig. \ref{fig: diag_sourcesonly}C we plot the peak-over-continuum strength of the 69~$\mu$m band versus the strength of the 11.3~$\mu$m band. This diagram is important because if the 11.3~$\mu$m and the 69~$\mu$m band come from the same population of crystalline olivine dust we expect these two bands to correlate with each other. The diagram indeed shows that the strength of the 69~$\mu$m band increases as the strength of the 11.3~$\mu$m band increases.

As a last diagram we also plot the strength of the 69~$\mu$m band against the width of the 69~$\mu$m band in fig. \ref{fig: diag_sourcesonly}D. This diagram can be used to investigate whether the abundance, that can be deduced from the 69~$\mu$m band, correlates with the grain temperature of crystalline olivine (which is a probe of the extend of the outflow). A weak anti-correlation can be seen between the peak-over-continuum strength and the width of the 69~$\mu$m band. This can be understood since a smaller width means colder crystalline olivine and a more strongly peaked 69~$\mu$m band. But we need to mention that the strength of the 69~$\mu$m band is not only depend on the temperature, but also on the abundance of the crystalline olivine. In sect. \ref{sec: models} we investigate all the different properties that have an effect on the strength of the 69~$\mu$m band

\begin{table}
\centering
\begin{tabular}{ l l}
  Dust species						&			Reference				\\
  \hline \hline
  \ Amorphous olivine ($\text{Mg}_{2}\text{SiO}_{4}$)	&			\cite{jager98}		\\	
  \ Amorphous olivine ($\text{MgFeSiO}_{4}$)	&			\cite{jager98}		\\
  \ Metallic iron (Fe (solid))				&			\cite{ordal88}		\\
  \ Crystalline olivine	($\text{Mg}_{2}\text{SiO}_{4}$)	&			\cite{suto06}		\\
\hline	
\end{tabular}
 \caption{Dust species used in this paper and references to the used optical constants}
\label{tab: refs}
\end{table}

\section{Modeling} 
\label{sec: models}
In this section we introduce a grid of models to interpret the observed crystalline olivine spectral features.
\subsection{Radiative transport and dust properties}
\label{sec: radtransport}
Model spectra were computed using the Monte Carlo radiative transport code MCMax \citep{min09}. Within the code we treat AGB stars as a central star enclosed in a spherically symmetric dust shell. The central star is treated as a black body and defined by its radius and temperature (see table \ref{tab: models}). We used a typical stellar temperature of 2700 K and luminosity of 7000 L$_{\odot}$ \citep{HO03}. This gives a radius for the central star of $380\,\text{R}_{\odot}\,=\,1.8\,\text{AU}$.

For the dust-to-gas ratio a value of 1/100 is assumed. The mass loss rate values mentioned in this paper always refer to the total (dust and gas) mass loss rates, assuming this dust-to-gas ratio. The temperature of the dust shell reaches a value of 1100 K at a radius of $\sim$ 8 AU. For all dust species this radius is taken to be the radius at which the dust species condense, and therefore defines the inner radius of the dust shell. By doing this we assume that the dust species formed before this 1100 K limit have a negligible effect on the spectrum. This is especially true for the optically thick dust shells where the infrared spectrum is not sensitive anymore to changes in the inner part of the dust shell. The terminal wind speed ($v_{\text{exp}}$) is set to the typical value of 10 km/s. In our modeling we use a density distribution proportional to $r^{-2}$, meaning that changes in the terminal wind speed only scale the mass loss rate values, but not change the model results.

The outflows of oxygen-rich AGB stars are mainly composed out of amorphous silicate dust. A moderate amount of crystalline silicates and metallic iron can be present. For the amorphous dust we use amorphous olivine ($(\text{Mg}, \text{Fe})_2\text{SiO}_4$). We use two laboratory measurement sets for the amorphous olivine: $\text{Mg}_{2}\text{SiO}_{4}$ and $\text{MgFeSiO}_{4}$ (see table \ref{tab: refs}). We can combine different ratios of these two amorphous olivines to simulate different amorphous olivine compositions. As a smooth continuum opacity we use metallic iron \citep{kemper02}. We also include crystalline olivine with zero percent iron. 

In our modeling we did not consider composite grains, such as grains that share a physical connection and thus their temperature. We do not consider composite grains because in these optically thick environments, the temperature of all dust species converges to the same value. The reason for different dust particles in the wind to reach the same temperature in an optically thick wind is that the absorption and emission of the grains happens at the same wavelength. In an optically thick wind the radiation field a dust particle can absorb is dominated by the infrared radiation of the surrounding dust grains. This contrasts to an optically thin wind, where absorption and emission can happen at different wavelengths and dust particles of different sizes or compositions can thus reach different temperatures. In such optically thin winds the radiation field that is absorbed by a dust grain is dominated by the radiation field of the central star, while the dust grain radiates at infrared wavelengths.

The optical constants used for these dust species are listed in table \ref{tab: refs}. We use the continuous distribution of ellipsoids \citep{BH83} shape distribution for the dust particles, which is valid for grains small compared to the wavelength of interest. The dust grains in the outflow of AGB stars are found not to exceed $\sim1\,\mu$m \citep{norris12}, making the CDE distribution applicable. We would like to note that the 69~$\mu$m band is independent of the grain size or shape distribution when small grains are considered ($<3.0\, \mu m$, \citealt{devries12, sturm13}).

The code MCMax handles temperature dependent opacities by using different opacity curves for the same dust species at different temperatures. For temperatures in between those temperatures for which opacity curves are available, MCMax interpolates between the two opacity curves. For temperatures above or below that for which opacity curves are available, MCMax uses the opacity curve for the highest or lowest temperature available, respectively. MCMax iterates to see if the opacities change the temperature structure in the shell and it updates the opacities at every iteration. 

In the outflow of OH/IR stars the temperature of the crystalline olivine grains can span a range from up to 1100~K at the condensation radius and down to 5~K or lower at the connection of the outflow to the ISM. For crystalline olivine (with no iron, x=0) optical constants are only available at 50, 100, 150, 200 and 295~K \citep{suto06}. From these data sets we know that at higher temperatures the 69~$\mu$m band will eventually be so weak and broad that it is indistinguishable from the continuum (see fig. \ref{fig: opacs_T}). Since the OH/IR stars we study contain at least some dust at these high temperatures, it could be important to take this into account. We have tested to see if it had an effect if we assume that at 600 K the crystalline olivine band at 69~$\mu$m has completely disappeared (while the other features are still the same as for 295K). We found that it did not matter for the properties of the crystalline olivine bands. The reason is that the crystalline olivine dust with temperatures much higher than 300 K are in the inner parts of the outflow (within the $\tau\sim1$ area) and therefore do not directly contribute to the spectrum. 

	\begin{figure}
	\begin{center}
	\resizebox{\hsize}{!}{\includegraphics{\absPath 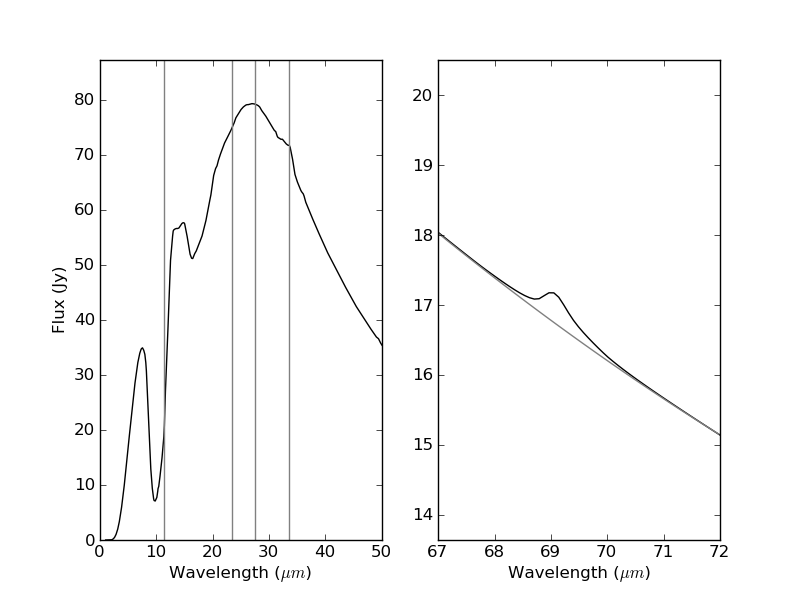}}
	\caption{Standard model spectra in the range of 2-50 $\mu$m and 67-72 $\mu$m. For parameters see table \ref{tab: models}. In gray the continuum under the 69~$\mu$m band. Spectral bands of crystalline olivine at 11.3, 23.5, 27.5 and 33.5 $\mu$m are indicated with vertical gray lines. The 23.5 and 27.5 $\mu$m bands are difficult to detect because these spectral bands are already in self reversal.}
	\label{fig: spec_std}
	\end{center}
	\end{figure}

\subsection{Outflow parameters}
We assume the star is currently in its superwind phase and has an enhanced mass loss reaching out to a radius of $R_{\text{SW}}$ from the central star. This means that this superwind started $R_{\text{SW}}/v_{\text{exp}}$ years ago, where $v_{\text{exp}}$ is the expansion velocity of the outflow. At radii outside $R_{\text{SW}}$ and up to $R_{\text{out}}$, an older outflow is present expelled before the superwind and at a mass loss rate a factor $\Phi$ lower than that of the superwind.  We take 500,000 AU as the outer radius because at this point the outflow has transitioned into the interstellar material. The dust shell reaches temperatures below 5~K at radii of 500,000 AU. Due to the interstellar radiation field the temperature of the outer radius of the dust shell could be significantly higher, for example \cite{li01} show that the temperature at the outer edge can be as high as 20 K or higher. Which outer radius is chosen is of no consequence to our results, since we show that this outer radius is of no influence on the properties of the 11.3~$\mu$m and 69~$\mu$m bands (see section \ref{sec: results}).

Within the superwind and in the pre-superwind mass loss we assume that the density distribution of our models is spherically symmetric and time independent with an $r^{-2}$ radial dependence. In our modeling approach we test the effect of the $R_{\text{SW}}$, $\Phi$, $R_{\text{out}}$ and the mass loss rate on the spectral features. From here on, mentioned mass loss rates always refer to that of the superwind. We also test the effects of the composition of the amorphous olivine, the metallic iron abundance and crystalline olivine abundance.

For the spatial distribution of the crystalline olivine we consider two cases. The first is that the crystalline olivine is only present in the superwind part of the outflow. This situation is expected, since crystalline material forms more efficiently in higher density regions  \citep{tielens98,gailsedl99,sogawa99, jones12}. In the second, the crystalline olivine is present in both the superwind and the pre-superwind mass loss. 

We also test if we can detect wether an OH/IR star has gone through one or several superwind phases. In our models the star is currently in a superwind phase, but we also test how the spectrum changes if another previously emitted superwind outflow is present. 

\begin{table}
\centering
\begin{tabular}{l l}
  Standard model parameters			&				 				\\

  \hline \hline
  mass loss rate ($10^{-5}\,\text{M}_{\odot}\,\text{yr}^{-1}$)	&			5							\\
  Inner radius (AU)					&				8							\\
  Superwind radius (R$_{\text{SW}}$, AU)		&				500 							\\
  Ratio of old to superwind mass loss  ($\Phi$)&			0.1							\\
  Outer radius dust shell (AU)			&				500,000						\\
  Luminosity central object (L$_{\odot}$)	&				7,000							\\
  Temperature central object (K)		&				2,700							\\
  Radius central object (R$_{\odot}$)	&				385						\\
  Terminal velocity ($v_{\text{exp}}$, km/s)	&				10							\\
  \hline
  \textit{Dust abundances (\% by mass):}	&										\\
  \ Amorphous olivine				&			92							\\
  \ Metallic iron						&			4 							\\
  \ Crystalline olivine					&			4 							\\
  \hline
  \textit{Dust composition:}				&										\\
  \ Amorphous olivine $\left(\frac{\text{Amount(Mg}_{2}\text{SiO}_{4})}{\text{Amount(MgFeSiO}_{4})} \right)$ &	0.5	\\
  \ Metallic iron						&			Fe (solid)						\\
  \ Crystalline olivine					&			$\text{Mg}_{2}\text{SiO}_{4}$				\\
  Dust shape model					&			CDE							\\
  Location of the crystalline olivine		&			$<R_{\text{SW}}$					\\
  \hline

\hline	
\end{tabular}
 \caption{Standard model parameters.}
\label{tab: models}
\end{table}

\subsection{Standard model}
\label{sec: standardmodel}
In order to visualize the effects of all the parameters we define a standard model for which we change one parameter at a time and investigate the changes (see table \ref{tab: models}). For our standard model we take a typical superwind with a radius of 500 AU and a mass loss rate of $5\cdot10^{-5}\,\text{M}_{\odot}\,\text{yr}^{-1}$. The mass loss rate observed for OH/IR stars ranges between $1\cdot10^{-5}\,\text{M}_{\odot}\,\text{yr}^{-1}$ and up to $15\cdot10^{-5}\,\text{M}_{\odot}\,\text{yr}^{-1}$ \citep{ches05, riechers05, just96, schut89, groene94}. 

It is not quite clear what the difference in mass loss rate is between the superwind and the pre-superwind phase. Modeling done in different studies on the same object (OH~26.5+0.6) resulted in different $\Phi$ values of 1/500 to 1/5 \citep{just96, ches05, groenewegen12}. \cite{just13} find a typical $\Phi$ value of the order $\sim0.01$ for the objects OH~127.8+0.0, WX~Psc, AFGL~5379, OH~26.5+0.6 and OH~30.1-0.7. As a standard value we take the old mass loss rate emitted before the superwind a factor of 10 ($\Phi=0.1$) lower than that of the superwind. We describe the effect of this parameter in sect. \ref{sec: results}. 

Crystalline olivine abundances in OH/IR stars can vary from less than 2\% to higher than 10\% \citep{devries10} and we choose a value of 4\% for our standard model. For the distribution of the crystalline olivine in the outflow we assume the crystalline olivine is only present in the superwind. Our standard model is currently in a superwind phase and in the standard model we do not include any previously emitted superwind. For the metallic iron abundance we take an abundance of 4\%, which is in the order of what \cite{kemper02} found for OH~127.8+0.0. The infrared spectrum of the standard model can be viewed in fig. \ref{fig: spec_std}. It is a model spectrum that displays the basic properties of observed OH/IR spectra in that it shows a clear 9.7 $\mu$m absorption band of amorphous olivine, that it peaks at wavelength longer than 20 $\mu$m and that it shows crystalline olivine bands at 33.6 and 69~$\mu$m.

\subsection{Modeling results}
\label{sec: results}
In this section we show the effects of different parameters on the infrared spectrum and its crystalline olivine spectral features. 

	\begin{figure}
	\begin{center}
	\resizebox{\hsize}{!}{\includegraphics{\absPath 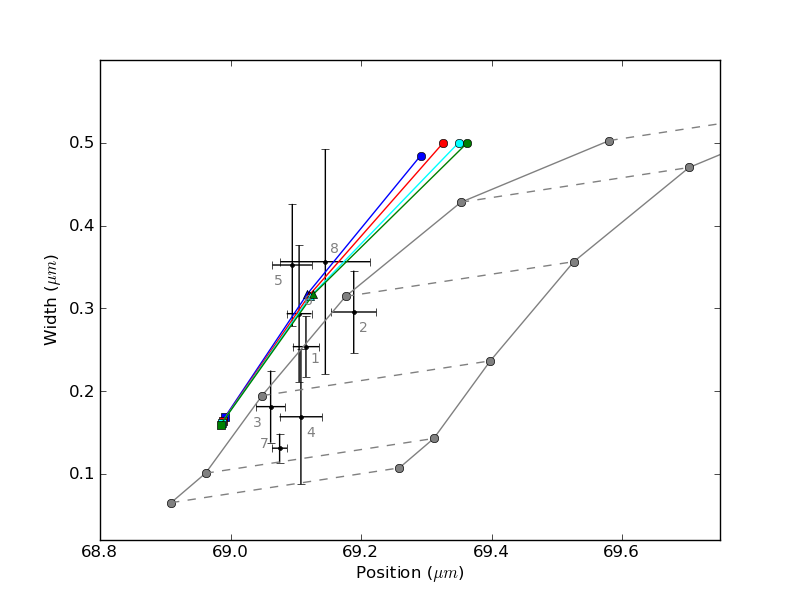}}
	\caption{The blue, red, teal and green curves are for model 69~$\mu$m bands with mass loss rates of $1\cdot10^{-5}\,\text{M}_{\odot}/\text{yr}$, $5\cdot10^{-5}\,\text{M}_{\odot}/\text{yr}$, $10\cdot10^{-5}\,\text{M}_{\odot}/\text{yr}$ and $15\cdot10^{-5}\,\text{M}_{\odot}/\text{yr}$, respectively. The circle, triangle and square are for superwind radii of 200, 500 and 2500 AU, respectively. All other model parameters are listed in table \ref{tab: models}. The width and position of the observed 69~$\mu$m bands are plotted with black dots. Gray solid and dashed lines are as in fig. \ref{fig: PW_sourcesonly}. 
	}
	\label{fig: MODEL_SPH_RSW_PW}
	\end{center}
	\end{figure}

	\begin{figure}
	\begin{center}
	\resizebox{\hsize}{!}{\includegraphics{\absPath 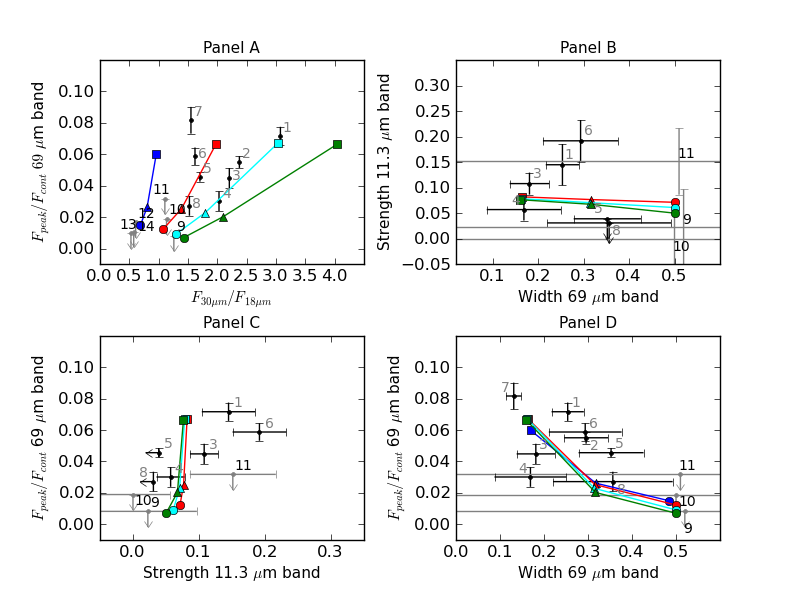}}
	\caption{The panels show different properties of model and observed spectra. The circle, triangle and square markers indicate models with \textbf{superwind radii} of 200, 500 and 2500 AU respectively. The blue, red, teal and green curves are for models with mass loss rates of $1\cdot10^{-5}\,\text{M}_{\odot}/\text{yr}$, $5\cdot10^{-5}\,\text{M}_{\odot}/\text{yr}$, $10\cdot10^{-5}\,\text{M}_{\odot}/\text{yr}$ or $15\cdot10^{-5}\,\text{M}_{\odot}/\text{yr}$ respectively. All other model parameters are listed in table \ref{tab: models}. The observations are shown as in fig. \ref{fig: diag_sourcesonly} .
	}
	\label{fig: MODEL_SPH_RSW_DIAG}
	\end{center}
	\end{figure}

\subsubsection*{\small{Superwind radius and mass loss rate}}
We test three radii for the superwind: 200, 500 and 2500 AU and we test these superwind radii at four different mass loss rates: $1\cdot10^{-5}\,\text{M}_{\odot}/\text{yr}$, $5\cdot10^{-5}\,\text{M}_{\odot}/\text{yr}$, $10\cdot10^{-5}\,\text{M}_{\odot}/\text{yr}$ and $15\cdot10^{-5}\,\text{M}_{\odot}/\text{yr}$. Fig. \ref{fig: MODEL_SPH_RSW_PW} shows that the position and width of the 69~$\mu$m band are not dependent on the mass loss rate for superwind radii larger than 200 AU. From fig. \ref{fig: MODEL_SPH_RSW_PW} it is also clear that less extended superwinds have crystalline olivine bands at 69~$\mu$m that are dominated by warmer grain temperatures than superwinds with larger radii. 

Figure \ref{fig:  MODEL_SPH_RSW_PW} shows that the width and position of model 69~$\mu$m bands are broader than those of the laboratory measurements. This is because there is a temperature gradient in the outflow, making the 69~$\mu$m band a superposition of 69~$\mu$m bands of different temperatures, effectively broadening the band. It further seems that the observations with narrower bands (\#3, 4 and 7) lay further away from the model bands than those with broader widths (among which \#5, 6 and 8). This could mean that the crystalline olivine in superwinds with larger radii contain slightly more iron.

Figure \ref{fig: MODEL_SPH_RSW_DIAG} shows the four panels with different spectral characteristics plotted against each other for models with varying superwind radii and mass loss rates. Panel A shows that the color of the spectrum becomes redder when the mass loss rate and/or the superwind radius increases. The peak-over-continuum strength of the 69~$\mu$m band is not strongly dependent on the mass loss rate, but is strongly influenced by the superwind radius. Panel B, C and D show that the strength of the 11.3~$\mu$m band is not dependent on the mass loss rate nor the superwind radius. The width of the 69~$\mu$m band on the other hand is a strong function of the superwind radius, but independent of the mass loss rate. 
	
\subsubsection*{\small{Crystalline olivine abundance}}
\begin{figure}
	\begin{center}
	\resizebox{\hsize}{!}{\includegraphics{\absPath 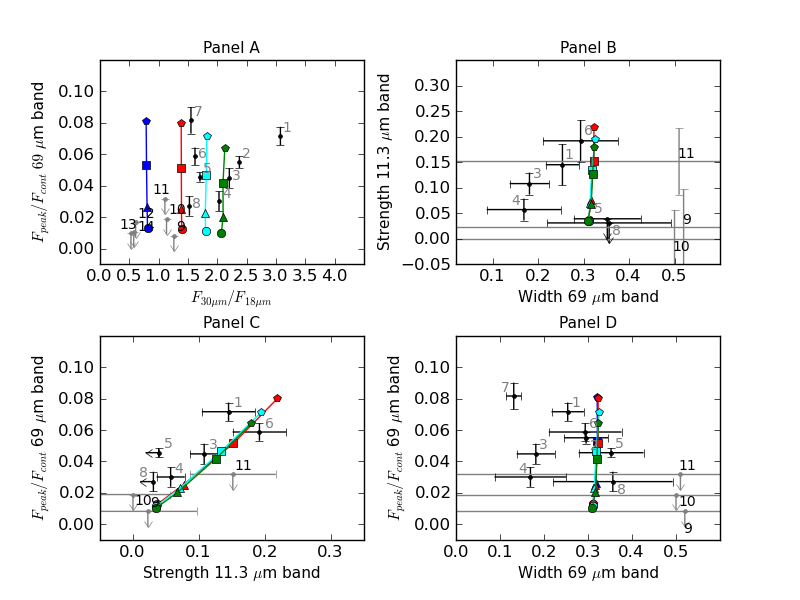}}
	\caption{The panels show different properties of model and observed spectra. The circle, triangle and square markers indicate models with \textbf{crystalline olivine abundances} of 2\%, 4\% and 8\% respectively. The other properties are as in fig. \ref{fig: MODEL_SPH_RSW_DIAG}.
	}
	\label{fig: MODEL_SPH_FO_DIAG}
	\end{center}
	\end{figure}

Figure \ref{fig: MODEL_SPH_FO_DIAG} shows models with different crystalline olivine abundances. It is clear that the abundance has no influence on the color of the spectrum or the width of the 69~$\mu$m band. What it does influence is the strength of the 11.3 and 69~$\mu$m bands. Because the crystalline olivine abundance has no effect on the width and position of the 69~$\mu$m band, no diagram like fig. \ref{fig: MODEL_SPH_RSW_PW} is shown for this parameter.

\subsubsection*{\small{Pre-superwind mass loss}}
\begin{figure}
	\begin{center}
	\resizebox{\hsize}{!}{\includegraphics{\absPath 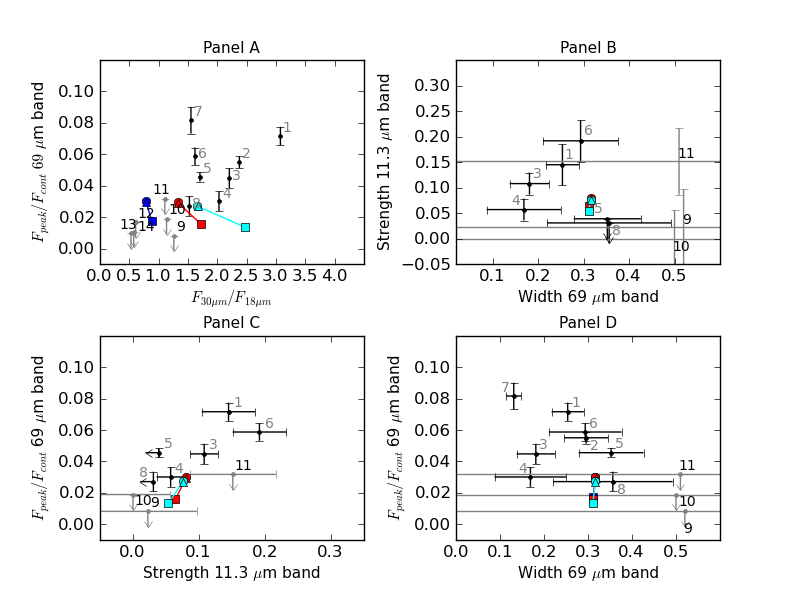}}
	\caption{The panels show different properties of model and observed spectra. The circle, triangle and square markers indicate models with \textbf{pre-superwind mass loss rate} fractions ($\Phi$) of 0.0, 0.01 and 0.5 respectively. The other properties are as in fig. \ref{fig: MODEL_SPH_RSW_DIAG}.
	}
	\label{fig: MODEL_SPH_FRACSW_DIAG}
	\end{center}
	\end{figure}

We have investigated the effect of the strength of the mass loss emitted prior to the superwind phase, by changing the parameter $\Phi$. By default $\Phi$ is set to 0.1, but here we also test the values 0.0 and 0.5. The $\Phi=0$ model corresponds to no mass loss prior to the superwind and $\Phi=0.5$ to a pre-superwind mass loss rate with a rate half that of the superwind. Note that in our standard model we do not include crystalline olivine in the pre-superwind mass loss phase. Figure~\ref{fig: MODEL_SPH_FRACSW_DIAG} shows that for $\Phi$ is 0.0 or 0.1 there is no significant change in any of the characteristics. In order to see any effect $\Phi$ has to be increased to 0.5. In this case a minor effect is seen on the color of the spectrum and the peak-over-continuum strength of the 69~$\mu$m band. The strength of this band decreases because the continuum flux increases when more mass is ejected during the pre-superwind period.

\subsubsection*{\small{Crystalline olivine distribution}}
\begin{figure}
	\begin{center}
	\resizebox{\hsize}{!}{\includegraphics{\absPath 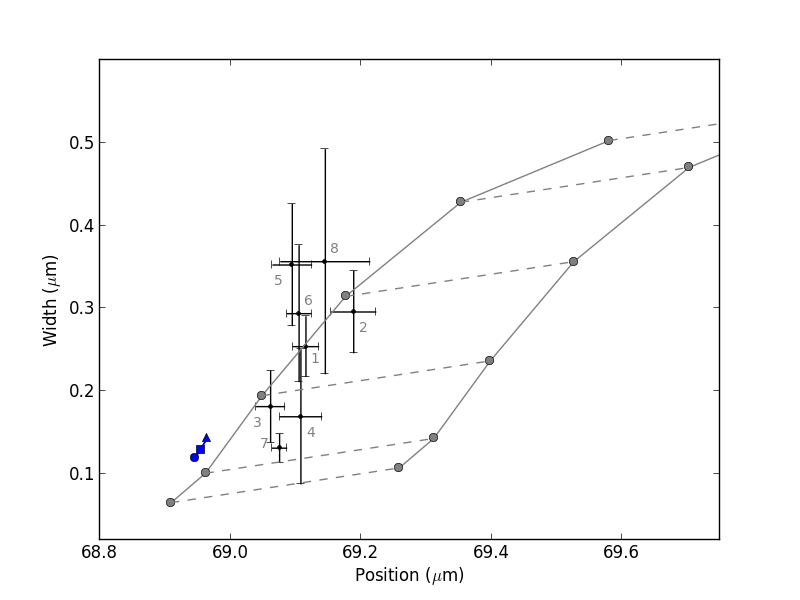}}
	\caption{The short blue curve connecting the circle, triangle and square in the lower left corner is for model 69~$\mu$m bands with a mass loss rate of $10\cdot10^{-5}\,\text{M}_{\odot}/\text{yr}$. We also calculated model bands for the mass loss rates $1\cdot10^{-5}\,\text{M}_{\odot}/\text{yr}$ and $5\cdot10^{-5}\,\text{M}_{\odot}/\text{yr}$, but these are not distinguishable from the $10\cdot10^{-5}\,\text{M}_{\odot}/\text{yr}$ models and they are not plotted. The circle, triangle and square are for superwind radii of 200, 500 and 2500 AU, respectively. In contrast to the standard model, in this case the crystalline olivine is also present in the pre-superwind mass loss period ejected before the superwind started. All other parameters are listed in table \ref{tab: models}. For the rest, the figure is the same as fig. \ref{fig: MODEL_SPH_RSW_PW}.
	}
	\label{fig: MODEL_SPH_FOINSW_PW}
	\end{center}
	\end{figure}

In our standard model we only have crystalline olivine present in the superwind part of the outflow, but in this section we investigate if we can indeed rule out the presence of crystalline olivine in the low mass loss rate outflow preceding the superwind phase. Figure \ref{fig: MODEL_SPH_FOINSW_PW} shows the position and width of the 69~$\mu$m band when crystalline olivine is present in both the superwind and pre-superwind mass loss period for three superwind radii. From this we can directly see that the 69~$\mu$m band is dominated by the cold grains far away from the star since the feature is very narrow. The feature corresponds to grain temperatures between 100 and 150 K. Because the feature is dominated by the coldest crystalline olivine grains, the feature is not sensitive anymore to the superwind radius.

The presence of crystalline olivine in the old mass loss part of the outflow does not affect the strength of the 11.3~$\mu$m band (not shown). It does affect the peak-over-continuum strength of the 69~$\mu$m, which is large compared to the observations with values $>$0.08 (not shown).

\subsubsection*{\small{Detectability of a previous superwind}}
In our standard model the star is currently in a superwind phase. If we assume that the superwind is linked to the thermal pulses of the star, then \cite{vassi93} show that for a 1 to 5 $\text{M}_{\odot}$ star the period between thermal pulses is 50,000-100,000 years. For an outflow with a velocity of 10 km/s this roughly computes to a distance between two superwind phases of 100,000-200,000 AU. For our model we take a current superwind radius of 500 AU and a previous superwind at 100,000 AU with a radial size corresponding to 10,000~AU. Crystalline olivine is present in both these superwind phases but not in the low mass loss rate periods in-between. 

\begin{figure}
	\begin{center}
	\resizebox{\hsize}{!}{\includegraphics{\absPath 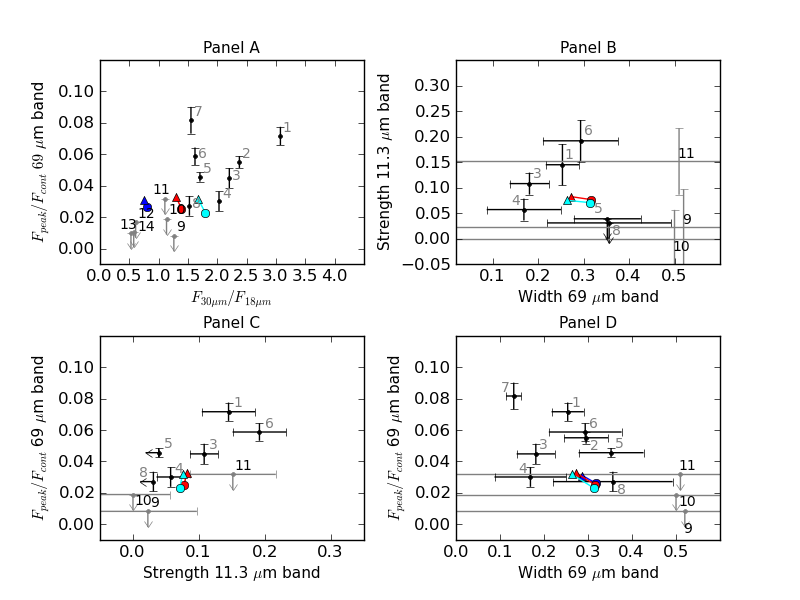}}
	\caption{The panels show different properties of model and observed spectra. The circle indicates a model with only one current superwind phase with a radius of 500 AU. The triangle indicates models that also have a \textbf{second superwind} at 100.000 AU of 10.000 AU long. The other properties are as in fig. \ref{fig: MODEL_SPH_RSW_DIAG}.
	}
	\label{fig: MODEL_SPH_SW2x_DIAG}
	\end{center}
	\end{figure}

Figure \ref{fig: MODEL_SPH_SW2x_DIAG} shows that the peak-over-continuum strength of the 69~$\mu$m slightly increases when a second superwind is present. The width of the 69~$\mu$m band becomes slightly narrower, because the old superwind adds emission of cold crystalline olivine grains to the spectrum, but this effect is insignificant compared to the effect of the superwind radius on the width of the 69~$\mu$m band. The strength of the 11.3~$\mu$m band is not influenced by a previously ejected superwind wind, since this band is only sensitive to the dust close to the central star.

\subsubsection*{\small{Metallic iron abundance}}
\begin{figure}
	\begin{center}
	\resizebox{\hsize}{!}{\includegraphics{\absPath 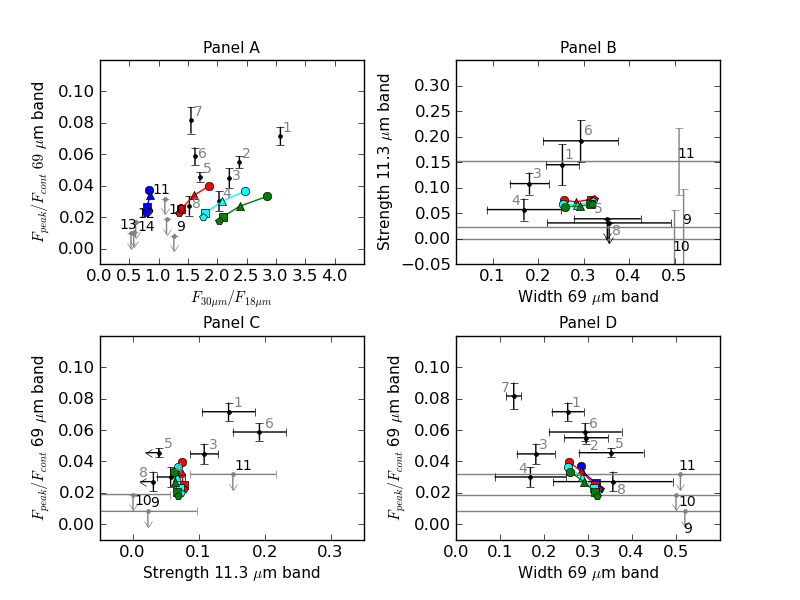}}
	\caption{The panels show different properties of model and observed spectra. The circle, triangle, square and pentagon markers indicate models with \textbf{metallic iron abundance} of 0\%, 1\%, 4\% and 8\%, respectively. The other properties are as in fig. \ref{fig: MODEL_SPH_RSW_DIAG}
	}
	\label{fig: MODEL_SPH_FE_DIAG}
	\end{center}
	\end{figure}

As a last parameter we also test the effect of the metallic iron abundance on the properties of the crystalline olivine bands. The outflow of the standard model contains 4\% metallic iron and we now also test the cases where the outflow contains 8\%, 1\% or 0\% iron. The results are shown in fig. \ref{fig: MODEL_SPH_FE_DIAG}. The metallic iron abundance does not have a significant effect on the strength of the 11.3~$\mu$m band. Metallic iron only has a minor effect on the color of the spectrum and the width and strength of the 69~$\mu$m band. It can be seen that less iron makes the spectrum redder, which is expected since metallic iron mostly adds opacity below $\sim$10 $\mu$m.

\subsubsection*{\small{Other parameters}}
Besides the parameters listed in this section we also tested the effects of the amorphous olivine composition and the outer radius of the total outflow. We tested the composition of amorphous olivine at the ratios $\left(\frac{\text{Amount(Mg}_{2}\text{SiO}_{4})}{\text{Amount(MgFeSiO}_{4})} \right)$ 0.2, 0.5 and 0.8. And we tested the outer radius at the values of 5,000 AU, 50,000~AU and 500,000 AU. It turned out that these two parameters have no effect on the general shape of the spectrum or the properties of the spectral bands of crystalline olivine and the corresponding diagrams are not shown.

\section{Results}
\label{sec: generalresults}

The parameter study of the previous sections shows that:
\begin{itemize}
\item The width of the 69~$\mu$m band is only dependent on the superwind radius
\item The strength of the 11.3~$\mu$m band is only dependent on the crystalline olivine abundance in the line of sight close to the star
\item The peak-over-continuum strength of the 69~$\mu$m band is only dependent on the superwind radius and the crystalline olivine abundance
\end{itemize}

\begin{figure}
	\begin{center}
	\resizebox{\hsize}{!}{\includegraphics{\absPath 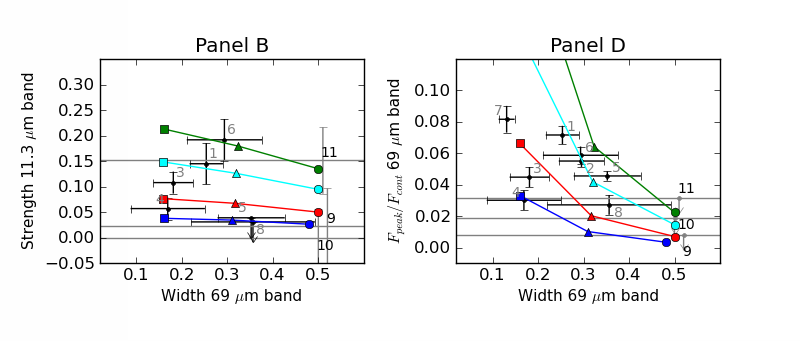}}
	\caption{The effect of the superwind radius and the crystalline olivine abundance on the 11.3 and 69~$\mu$m band for models with a mass loss rate of $15\cdot10^{-5}\,\text{M}_{\odot}/\text{yr}$. The circle, triangle and square markers indicate models with \textbf{superwind radii} of 200, 500 and 2500 AU respectively. The blue, red, teal and green curves now represent the \textbf{crystalline olivine abundances} of 2, 4, 8 and 12 \%. The other parameters of the models are as in the standard mode. The other properties are as in fig. \ref{fig: MODEL_SPH_RSW_DIAG}.
	}
	\label{fig: MODEL_SUM_DIAG_15}
	\end{center}
	\end{figure}

Panel B of fig. \ref{fig: MODEL_SUM_DIAG_15} shows the effect of the superwind radius and abundance on the width of the 69~$\mu$m band and the strength of the 11.3~$\mu$m band for the mass loss rate $15\cdot10^{-5}\,\text{M}_{\odot}/\text{yr}$. We can now use fig. \ref{fig: MODEL_SUM_DIAG_15}B to read off the superwind radius and crystalline olivine abundance in the line of sight. We listed these values in table \ref{tab: tableDisc}. The crystalline olivine abundance in the line of sight varies from lower than 2\% to higher than 10\%, which is comparable to that found by \cite{devries10} using the same method. The superwind radii of the sources with 69~$\mu$m bands can be as small as 200 AU and go up to $\sim$2500 AU. 

The abundance in the line of sight could not be determined for the sources with 9.7 $\mu$m bands in-between an emission and absorption band (RAFGL~2374, WX~Psc and IRC~+50137). And for the sources with no detected 69~$\mu$m bands (OH~104.91+2.4, AFGL~4259, IRAS~17010, RAFGL~2374, WX~Psc and IRC~+50137) we can only determine an upper limit of the size of the superwind radius. We argue that these six sources have no 69~$\mu$m bands because they only contain very hot ($>$300 K) crystalline olivine, since the opacities of such a population of hot crystalline olivine grains do not show a feature at 69~$\mu$m even though they show a feature at 33.6 $\mu$m. In order to have an outflow with no cold and only hot crystalline olivine grains, we suggest that these sources have superwind radii smaller than 200 AU. Such a small superwind radius is also consistent with the fact that these sources have bluer infrared spectra compared to the other sources with larger superwind radii. Indeed \cite{decin07} showed that the current mass loss phase of WX~Psc only extends up to $\sim$50 AU.

Now that we know the superwind radii of our sources with 69~$\mu$m bands, we can also derive a crystalline olivine abundance from the peak-over-continuum strength of the 69~$\mu$m band. Panel~D from fig. \ref{fig: MODEL_SUM_DIAG_15} shows the effect of the superwind radius and crystalline olivine abundance on the peak-over-continuum strength of the 69~$\mu$m band. Table \ref{tab: tableDisc} lists the abundances derived from the 69~$\mu$m band. 

In the case the crystalline olivine abundance does not spatially vary in the outflow, both the abundance derived from the 11.3~$\mu$m and the 69~$\mu$m band must agree with each other. For the majority of the sources the abundance in the line of sight is indeed similar to that derived from the 69~$\mu$m band, but a significant deviation is found for OH~127.8+0.0. For OH~127.8+0.0 the abundance in the line of sight is $<$2\% while that derived from the 69~$\mu$m band is $9\pm2$\%. That the abundance from the 11.3~$\mu$m and the 69~$\mu$m band agree for most of our sources, indicates that indeed the crystalline olivine is part of the outflow and not, for example, likely to be abundant in a stable disk-like structure embedded in the outflow.

The abundance mismatch found for OH~127.8+0.0 can either be due to a radial abundance gradient or a spherical abundance gradient in the outflow. A lower abundance from the 11.3~$\mu$m band can be explained by having a lower abundance close to the star compared to farther away from the star. But since the superwind of OH~127.8+0.0 is only 400$\pm$200~AU, this is not very plausible. It is more likely that the abundance of crystalline olivine is higher in the equatorial plane and that the equatorial plane is seen pole on. In that case the 69~$\mu$m band would probe the full equatorial plane and result in a higher abundance than the 11.3~$\mu$m band which only probes the line of sight towards the pole. 

In the modeling section we also presented two scenarios where crystalline olivine is only present in the superwind or when it is also present in the mass loss period emitted before the superwind. When the crystalline olivine is also present in the pre-superwind mass loss period (with a mass loss rate of $>$10\% that of the superwind phase) the 69~$\mu$m bands are so narrow and so strong that they do not compare well with our observed sources (see fig. \ref{fig: MODEL_SPH_FOINSW_PW}). This shows that the crystalline olivine is only present in the superwind part of the outflow. 

We also tested the presence of a previously emitted superwind containing crystalline olivine. From fig. \ref{fig: MODEL_SPH_SW2x_DIAG} it can be seen that the effect of a previously emitted superwind on the 11.3~$\mu$m and 69~$\mu$m band is small and that we can therefore not rule out the presence of such a previously emitted superwind. 

The composition of the crystalline olivine in the outflow of these OH/IR stars is discussed by Blommaert et al. (in prep.). They show, using fig. \ref{fig: PW_sourcesonly}, that the crystalline olivine contains almost no iron (0-0.5\%). If we compare the width and position of the 69~$\mu$m bands with the models with different superwind radii in fig. \ref{fig: MODEL_SPH_RSW_PW}, we might see a slight difference in iron content as a function of superwind radius. The OH/IR stars with the smallest superwind radii (for example AFGL~5379 and OH~127.8+0.0) correspond very well with our models using pure forsterite (no iron). On the other hand, the sources with larger superwind radii (for example IRAS~21554 and OH 26.5+0.6) have slightly red shifted 69~$\mu$m bands and the crystalline olivine might contain a small amount of iron ($<$0.5\%).

\beginthetable
\centering
\begin{tabular}{ l l l l l l l}
  Source name	&	\#	&11.3~$\mu$m	&69~$\mu$m 	&$R_{\text{SW}}$ (AU)		&$A_{11.3}$ (\%)		&$A_{69}$ (\%)\\
  \hline \hline
OH~32.8-0.3  	&	1	&+			&+			&1,500$\pm$400	&8$\pm$2				&7$\pm$2			\\%
OH~21.5+0.5 	&	2	&?			&+			&1,250$\pm$250	&?					&7$\pm$2			\\
OH~30.1-0.7  	&	3	&+			&+			&2,500$\pm$300	&5$\pm$2				&3$\pm$1			\\%
OH~26.5+0.6  	&	4	&+			&+			&2,500$\pm$1,000	&3$\pm$1				&2$\pm$1			\\%
OH~127.8+0.0 	&	5	&-			&+			&400$\pm$200		&{$<$2}				&{9$\pm$2}	\\
AFGL~2403 	&	6	&+			&+			&700$\pm$400		&11$\pm$3			&8$\pm$2			\\%
IRAS~21554	&	7	&?			&+			&2,800$\pm$300	&?		 			&4$\pm$2			\\
AFGL~5379 	&	8	&-			&+			&400$^{+1000}_{-200}$&$<$2				&5$\pm$3			\\%
OH~104.91+2.4&	9	&-			&-			&$<$200			&$<$2				&$<$2			\\%
AFGL~4259 	&	10	&-			&-			&$<$200			&$<$2				&$<$2			\\%
IRAS~17010	&	11	&+			&-			&$<$200			&10$\pm$4			&$<$12			\\%
RAFGL~2374	&	12	&?			&-			&$<$200			&?					&.				\\%
WX~Psc  		&	13	&?			&-			&$<$200			&?					&.				\\%
IRC~+50137	&	14	&?			&-			&$<$200			&?					&.				\\%
\hline	
\end{tabular}
 \caption{List of the observed OH/IR stars and modeling results. The two columns 11.3~$\mu$m and 69~$\mu$m list if these bands of crystalline olivine are detected. The next columns lists the superwind radii found from the 69~$\mu$m band, the crystalline olivine abundance derived from the 11.3 and that derived from the 69~$\mu$m band.}
\label{tab: tableDisc}
\endthetable

\section{Discussion and Conclusions}
\label{sec: discussion}
Since the crystalline olivine is part of the superwind, we can take the crystalline olivine as a tracer of the bulk of the dust in the superwind. This means that the width of the 69~$\mu$m bands shown in fig.~\ref{fig: MODEL_SPH_RSW_PW} can only be explained if the superwind radii of these stars is smaller than $\sim$2500~AU = 1400~R$_{*}$. A maximum superwind radius of 2500~AU means, assuming an outflow velocity of 10-15~km/s, that the outflow started roughly $\lesssim$1200-800~years ago. 

Recently more evidence has been found that indeed the duration of the superwind is very short. The study, based on SED modeling and OH and CO observations, of \cite{just06} showed for several stars (among others OH~30.1-0.7, OH~26.5+0.6, OH~21.5+0.5, OH~32.8-0.3 and AFGL~5379) that the superwind must have started $<$2000~years ago. 

Based on CO observations \cite{just13} find superwind durations shorter than 300~years, for five objects that are also in our sample (WX~Psc, OH~127.8+0.0, AFGL~5379, OH~26.5+0.6 and OH~30.1-0.7). The $<$300~years found by \cite{just13} corresponds to a superwind radius of $<$600~AU. In the case of WX~Psc, OH~127.8+0.0 and AFGL~5379 this corresponds well with our findings (see table \ref{tab: tableDisc}), but the superwind radius we find for OH~26.5+0.6 and OH~30.1-0.7 is larger than that of \cite{just13}. Independent SED modeling of OH~26.5+0.6 \citep{groenewegen12, ches05} resulted in a superwind radius for this object of $\sim$500 AU, which is smaller than the 2500$\pm$1000~AU that we find.

The mass loss rates of OH/IR stars can go up to ${15\cdot10^{-5}\,\text{M}_{\odot}/\text{yr}}$ and even slightly higher \citep{just96, just92, schut89, groene94}. If we combine these mass loss rates with the timescales we derive, we find a total mass in the superwind of $\sim$0.2~$\text{M}_{\odot}$ (at a gas over dust ratio of 100). The gas over dust ratio can be as high as 200-300 \citep{lombaert13}, meaning the star could lose 0.2-0.6~$\text{M}_{\odot}$ in one superwind phase. Depending on the initial mass of the star, one superwind could already contain enough mass for the AGB star to evolve away from the AGB. 

One object that is known to originate from an OH/IR star and that just ascended the post-AGB is HD 161796. This object has been extensively analyzed by \cite{hoogzaad02} and \cite{min13} and is shown to have stopped its mass loss only $\sim$300 years ago. HD 161796 now has an expanding oxygen-rich dust shell that contains $\sim$0.05 $\text{M}_{\odot}$ of gas and dust (\citealt{min13} and private communication). This mass is consistent with the mass we find that an OH/IR stars can lose in one superwind phase. The mass of HD 161796 before it ascended the AGB is found to be $\sim1\,\text{M}_{\odot}$ \citep{kipper07, stasinska06}, making this object an example of a star with a low mass that left its AGB phase with only one superwind phase. In the case of OH/IR stars with higher progenitor masses, the star would need to go through several superwind phases before it could leave the AGB phase. 

There are strong indications that the OH/IR stars in our sample have high initial masses ($\geq$5 M$_{\odot}$). First of all most of our objects have long periods (order of $\sim$1000~days) and/or high luminosities ($>$10,000 L$_{\odot}$) and they are situated at low galactic latitudes (\citealt{vanlangevelde90} and \citealt{beck10} and references therein, see table \ref{tab: tableSources}). Furthermore, based on the $^{12}$C/$^{13}$C and $^{18}$O/$^{17}$O ratio it is shown that most of our sources have experienced hot-bottom burning \citep{delfosse97,just13}, indicating that they must have a high initial mass ($\geq$5 M$_{\odot}$). OH/IR stars with such high progenitor masses need to lose several solar masses before they can leave the AGB. Therefore we expect to see several superwind phases as extended shells around the OH/IR stars in our sample, in the same way as these extended shells are seen around carbon-rich AGB stars \citep{cox12, maercker12}. 

Using the methods in this paper we can not study any previously emitted superwinds because the width and strength of the 69~$\mu$m band are not sensitive to the presence of previously emitted superwind phases. But it is intriguing that even though several oxygen-rich AGB stars have been observed with Herschel, none of them show any extended structure \citep{cox12}. If this is an indication of the fact that these OH/IR stars indeed have no extended shells and thus no previously emitted superwinds, then it is hard to understand how these OH/IR stars lose enough mass to evolve away from the AGB.

The next step in understanding the evolution of these OH/IR stars is to systematically search for extended material around these stars. This would show if these objects have gone through one or several superwind phases. If it is indeed the case that OH/IR stars only have one superwind, than these stars need to lose their mass in another way. We hypothesize that these OH/IR stars evolve into an as of yet unrecognized type of object with an even more extremely high and abrupt mass loss phase. Such an object would have an SED so extremely red it would not necessarily be recognized as an OH/IR star. 

The post-AGB or proto-PN IRAS 16342$-$3814 could be an example of such an object, since its SED is extremely red. The SED shows no flux at wavelengths at and below 10 $\mu$m and peaks around $\sim$40 $\mu$m \citep{dijkstra03}. The crystalline olivine bands seen in the spectrum of IRAS 16342$-$3814 are also in absorption up to 45 $\mu$m. The object has an almost spherical dust shell that is punctured by bipolar jets \citep{verhoelst09}. The mass loss rate it must have had on the AGB is of the order of $10^{-3}\,\text{M}_{\odot}/\text{yr}$ \citep{verhoelst09}. Maybe OH/IR stars with massive progenitors evolve into a phase like that of IRAS~16342$-$3814, in which they develope an even higher mass loss rate for a short duration after which they directly evolve away from the AGB.

\begin{acknowledgements} 
   B.L. de Vries acknowledges support from the Fund for Scientific Research of
   Flanders (FWO) for his Aspirant fellowship as well as under grant number G.0470.07. 
\end{acknowledgements}

\bibliographystyle{aa}
\bibliography{references}

\end{document}